\documentclass[]{emulateapj}

\usepackage{natbib}
\def\eg{e.g.}

\def\ie{i.e.}

\def\etc{{\it etc.}}

\def\about{$\sim$}
\def\simlt{\buildrel{<}\over \sim}
\def\simgt{\buildrel{>}\over \sim}

\def\simlt{$\la$}
\def\simgt{$\ga$}

\def\asec{$''$}
\def\amin{$'$}

\def\yr-1{yr$^{-1}$}

\def\Chandra{{\it Chandra}}

\def\bestz{{\it bestz}}
\def\photoz{$Photoz$}
\def\photoz2{$Photoz2$}

\begin{document}

\title{The Field X-ray AGN Fraction to $\lowercase{z}=0.7$ from the {\it
Chandra} Multiwavelength Project and the Sloan Digital Sky
Survey}

\author{Daryl Haggard\altaffilmark{1,2,3}, 
        Paul J. Green\altaffilmark{4},
        Scott F. Anderson\altaffilmark{1},
        Anca Constantin\altaffilmark{5},
        Tom L. Aldcroft\altaffilmark{4},
        Dong-Woo Kim\altaffilmark{4}, 
        Wayne A. Barkhouse\altaffilmark{6}}
	
\altaffiltext{1}{Department of Astronomy, University of Washington,
  Box 351580, Seattle, WA 98195; dhaggard@astro.washington.edu}
\altaffiltext{2}{NASA Harriett G. Jenkins Fellow}
\altaffiltext{3}{Kavli Institute for Theoretical Physics, 
  University of California, Santa Barbara, CA 93106, USA}
\altaffiltext{4}{Harvard-Smithsonian Center for Astrophysics, 60
  Garden Street, Cambridge, MA 02138, USA}
\altaffiltext{5}{Department of Physics and Astronomy, 
  James Madison University, Harrisonburg, VA 22807, USA}
\altaffiltext{6}{Department of Physics and Astrophysics, University of
  North Dakota, Grand Forks, ND 58202, USA}

\begin{abstract}

We employ the \Chandra\ Multiwavelength Project (ChaMP) and the Sloan
Digital Sky Survey (SDSS) to study the fraction of X-ray-active
galaxies in the field to $z = 0.7$. We utilize spectroscopic redshifts
from SDSS and ChaMP, as well as photometric redshifts from several
SDSS catalogs, to compile a parent sample of more than 100,000 SDSS
galaxies and nearly 1,600 \Chandra\ X-ray detections. Detailed ChaMP
volume completeness maps allow us to investigate the local fraction of
active galactic nuclei (AGN), defined as those objects having
broad-band X-ray luminosities $L_X$(0.5-8~keV)~$ \ge
10^{42}$~erg\,s$^{-1}$, as a function of absolute optical magnitude,
X-ray luminosity, redshift, mass, and host color/morphological
type. In five independent samples complete in redshift and $i$-band
absolute magnitude, we determine the field AGN fraction to be between
$0.16\pm{0.06}$\% (for $z \le 0.125$ and $-18 > M_i > -20$) and
$3.80\pm{0.92}$\% (for $z \le 0.7$ and $M_i < -23$). We find excellent
agreement between our ChaMP/SDSS field AGN fraction and the \Chandra\
cluster AGN fraction, for samples restricted to similar redshift and
absolute magnitude ranges: $1.19\pm{0.11}$\% of ChaMP/SDSS field
galaxies with $0.05 < z < 0.31$ and absolute $R$-band magnitude more
luminous than $M_R < -20$ are AGN. Our results are also broadly
consistent with measures of the field AGN fraction in narrow, deep
fields, though differences in the optical selection criteria, redshift
coverage, and possible cosmic variance between fields introduce larger
uncertainties in these comparisons.

\end{abstract}


\section{Introduction}
\label{intro}

A fundamental constraint on all theories modeling the interplay of
supermassive black hole (SMBH) accretion and galaxy evolution should
be the fraction of galaxies in the local universe that host actively
accreting nuclei. Accretion onto a SMBH is the predominant source of
energy produced by active galactic nuclei (AGN) and QSOs, so a
measure of accretion activity (at a variety of redshifts and
luminosities) provides a valuable constraint on the black hole mass
function at the present day \citep{Soltan82,Rees84,Marconi04}.

In both active and inactive galaxies, observations show a remarkably
tight correlation between the properties of galactic spheroids and the
masses of their central black holes, the $M_{BH}-\sigma$ relation
\citep{Ferrarese00,Gebhardt00,Gultekin09}, suggesting that SMBHs play
an important role in the formation of galactic bulges. Simulations
indicate that the black hole and the bulge may co-evolve, each growing
as a result of repeated merger events
\citep[\eg,][]{Hernquist89,DiMatteo05,Hopkins05,Hopkins08}. In this
scenario, the merger causes cold gas to fall to the center, triggering
starbursts, and fueling rapid black hole growth. Even while accretion
inflows increase the black hole's mass, outflows transport energy from
the active nucleus to the surrounding host galaxy
\citep[\eg,][]{Granato04}, creating a mechanism by which the galaxy
and it's central black hole can ``feedback'' as they evolve. In
extreme cases, these outflows can carry enough energy to alter the
galaxy's evolution by ejecting gas and truncating star formation.

Several well-studied phenomena support this so-called feedback
paradigm, which links star formation and SMBH growth to galaxy
mergers. First, the space density of low-luminosity active galactic
nuclei (LLAGN) peaks at lower host mass than that of luminous quasars
\citep{Barger05a,Hasinger05,Scannapieco05}, indicating that AGN may
undergo ``cosmic downsizing'' as do galaxy spheroids
\citep{Cowie96}. Next, the local mass density of SMBHs corresponds
closely to the luminosity density produced by quasars at high redshift
\citep{Yu02,Hopkins06}. Hence, the present-day population of AGN and
quiescent black holes may represent the fossil record of their
brighter, high-redshift counterparts (the QSO luminosity function
peaks at redshifts $z \sim 2$), formed at a time when galaxy mergers
were common. This suggests that all massive galaxies may have
undergone an AGN phase \citep[\eg, ][]{Marconi04,Shankar04}.

Though they clearly influence the evolution of galaxies and the
conditions in the early universe, fundamental questions about the
formation and evolution of accreting black holes remain. To be robust,
the merger and feedback paradigm must reproduce the fraction of
galaxies hosting AGN, the ratio of obscured to unobscured AGN
necessary to match the cosmic X-ray background, the space density of
SMBHs, and AGN clustering.

X-ray emission is the most reliable primary signature of AGN activity
because X-rays offer the most complete and efficient marker of active
accretion close to the black hole (10--100 gravitational
radii). Combined with other wavelengths, X-ray observations sample a
wide variety of AGNs and improve our census of AGN demographics
\citep[][and references therein]{Brandt05}.

Galaxies that do not harbor a powerful accreting black hole may also
produce X-rays due to hot interstellar gas, supernova remnants, and
low- and high-mass X-ray binary populations \citep{Fabbiano06,Kim06};
and even in ``normal'' galaxies a LLAGN may persist \citep[see][for a
review]{Ho08}. In this work, we consider the X-ray-active fraction of
galaxies out to $z = 0.7$, in a field covering more than $20~deg^2$,
at a wide range of broad-band X-ray luminosities ($10^{40}$ \simlt\
$L_X$[0.5-8~keV;~erg\,s$^{-1}$] \simlt\ $10^{46}$), including those
characteristic of normal and starforming galaxies. Since the many
X-ray-producing mechanisms described above can muddy the waters in
lower-X-ray-luminosity systems, we focus the bulk of our analysis on
sources with X-ray luminosities $L_X$(0.5-8~keV)~$ \ge
10^{42}$~erg\,s$^{-1}$, characteristic of accreting SMBHs.

Measurements of the AGN fraction as a function of environment have
been made using both optical spectroscopic samples and
optical/radio-selected samples. \cite{Kauffmann04} use emission-line
galaxies in the Sloan Digital Sky Survey \citep[SDSS;][]{York00} to
measure the fraction as a function of environment and find that the
fraction of very luminous AGN decreases with increasing environmental
density \cite[see also][]{Popesso06}; while less luminous SDSS AGN
show no such trend \citep{Miller03,Kauffmann04}. Studies of the
radio-loud SDSS AGN fraction demonstrate that the fraction of galaxies
hosting radio-loud AGN increases towards richer environments
\citep[except at the highest emission-line luminosities,][]{Best05}.

As in field studies, X-ray identification of AGN activity in clusters
has proved to be even more efficient than optical or radio techniques,
particularly for low-luminosity AGN. A significant body of work on the
cluster AGN fraction has been enabled by a combination of \Chandra\
X-ray observations and spectroscopic follow-up for a series of 32
clusters with $0.05 < z < 1.3$
\citep{Martini02,Martini06,Martini07,Martini09}. These authors find
larger AGN fractions in clusters than those found in optical emission
line studies. One of the aims of the present work is to make a
detailed comparison between the field and cluster X-ray AGN fractions
based on a combination of \Chandra\ imaging and ground-based optical
photometry.

NASA's {\it Chandra X-ray Observatory} \citep{Weisskopf02} has
generated a rich archive of X-ray sources, detected with unprecedented
resolution and sensitivity. By cross-correlating the \Chandra\ archive
with the SDSS --- an extensive optical survey in five filters --- it
is possible to identify X-ray counterparts to individual optical
objects, confirming their classification as quasars or active galactic
nuclei. The \Chandra\ Multiwavelength Project
\citep[ChaMP;][]{Green04,Green09} has carefully analyzed 323 \Chandra\
fields (about 30 square degrees) that overlap the SDSS and
characterized all optical/X-ray matches.  We employ these X-ray and
optical surveys to determine the fraction of actively accreting SMBHs
(or AGN) out to redshift of $\sim 0.7$. We also explore the AGN
fraction as a function of galaxy luminosity, redshift, mass, and host
galaxy morphology. Such measures of the AGN fraction will prove useful
for constraining the fueling, lifetimes, and growth of central SMBHs.

In \S \ref{catalogs} we describe the Extended ChaMP (\Chandra\ Cycles
1--6) and SDSS surveys; in \S \ref{sample_select} we detail the
ChaMP/SDSS sample developed for this study, including specifics on the
spectroscopic and photometric redshift catalogs we employ. Section
\ref{properties} outlines our methods for determining k-corrections,
absolute magnitudes, masses, X-ray and optical luminosities, and our
definition of five independent samples complete in redshift and
$i$-band absolute magnitude. In \S \ref{the_frac} we describe our
determination of the X-ray-active and AGN fractions in these
volume-limited optical samples, investigate possible evolution of the
fraction with redshift, and trends with absolute magnitude and mass,
in addition to providing an assessment of how photometric redshift
errors and other selection effects might impact our measure of the AGN
fraction. We discuss the AGN fraction as a function of galaxy color
and compare our results to others in the field, groups, and clusters
in \S \ref{discussion}; and wrap up with our conclusions in \S
\ref{conclusions}.

We adopt an $H_0 = 70$~km\,s$^{-1}$\,Mpc$^{-1}$, $\Omega_{\Lambda} =
0.7$, and $\Omega_M = 0.3$ cosmology throughout.

\section{The X-ray and Optical Catalogs}
\label{catalogs}

\subsection{The Extended \Chandra\ Multiwavelength Project}
\label{champ}

The \Chandra\ Multiwavelength Project is a wide-area serendipitous
X-ray survey based on archival X-ray images of the $|b| > 20~deg$ sky
observed with the AXAF CCD Spectrometer \citep[ACIS;][]{Weisskopf02}
on-board \Chandra \footnote{ChaMP results and data are available
on-line: http://hea-www.harvard.edu/CHAMP.}. The full 130-field Cycle
1--2 X-ray catalog is public \citep{Kim07b}, and the most
comprehensive X-ray number counts (log $N$-log $S$) to date have been
produced thanks to 6600 sources and massive source-retrieval
simulations \citep{Kim07a}.

We have recently expanded our X-ray analysis to cover a total of 392
fields through \Chandra\ Cycle 6 \citep[see detailed descriptions
in][]{Covey08,Green09}, to improve statistics and encompass a wider
range of source types --- we refer to this expansion as the Extended
ChaMP. The new list of \Chandra\ pointings avoids overlapping X-ray
observations by eliminating the observation with the shorter exposure
time. As described in \cite{Green04}, we also avoid \Chandra\ fields
with large (\simgt 3 \amin) extended sources in either optical or
X-rays (\eg, nearby galaxies M101, NGC 4725, NGC 4457, or clusters of
galaxies MKW8, or Abell 1240). Spurious X-ray sources (due to, \eg,
hot pixels, bad bias, bright source readout streaks) have been flagged
and removed as described in \cite{Kim07b}. For the expansion, we
select only \Chandra\ fields that overlap SDSS Data Release 5 imaging;
of the 392 ChaMP fields (observation IDs; hereafter ``obsids''), 323
fall within the SDSS DR5 footprint\footnote{A list of the \Chandra\
fields included in this study (obsid, R.A., Dec, exposure time [ksec],
number of ChaMP/SDSS matches, and total number of optical spectra) can
be found at http://hea-www.cfa.harvard.edu/CHAMP/.}. The Extended
ChaMP covers \about $30~deg^2$ at the brightest fluxes\footnote{The
brightest ChaMP fluxes are $\sim f_x (0.5-8.0 {\rm keV}) \sim 5 \times
10^{-13}$~erg\,cm$^{-2}$\,s$^{-1}$, where the flux limit represents
the number of counts detectable in 90\% of simulation trials,
converted to flux assuming a power-law $\Gamma = 1.7$ at $z = 0$ and
the Galactic $N_H$ appropriate for each obsid.}  \citep{Green09}. In
addition to SDSS imaging, we have undertaken deep ($r \sim 25$)
NOAO/MOSAIC optical imagining \citep[described in][]{Green04}, work
currently being extended to 67 fields (W.A. Barkhouse et al. 2010, in
preparation).

We have incorporated spectroscopic information, redshifts and line
diagnostics, for the entire optical catalog from: (1) existing ChaMP
spectroscopy \citep{Green04}; (2) the SDSS DR7 spectroscopic catalog
\citep{Abazajian09}; (3) the SDSS DR7 Max Planck Institute for
Astrophysics/John's Hopkins University (MPA/JHU) value-added galaxy
catalog \citep{Brinchmann04}\footnote{Publicly available at
http://www.mpa-garching.mpg.de/SDSS/.}; and (4) cross-correlation with
the NASA Extragalactic Database (NED). \cite{Green09} and
\cite{Constantin09a} give details on matching the optical spectroscopy
with the Extended ChaMP X-ray source catalog.

The Extended ChaMP also includes a comprehensive set of sensitivity
maps for ACIS imaging, implemented in the {\it xskycover} table. This
allows (1) recognition of imaged-but-undetected objects, (2) counts
limits for 50\% and 90\% detection completeness, (3) corresponding
flux upper limits at any sky position, as well as (4) flux sensitivity
versus sky coverage for any subset of obsids. To generate the table we
use the {\it wavdetect} detection algorithm in CIAO \citep{Freeman02}
to generate threshold maps at each {\it wavdetect} kernel scale (1, 2,
4, 8, 16, and 32 pixels). The threshold maps, computed from the local
background intensity, determine the magnitude of the source counts
necessary for a detection at each pixel with a detection threshold of
$P = 10^{-6}$ (corresponding to 1 false source per $10^6$
pixels). Thus, when an object is not detected at a given location, the
threshold map value serves as an upper limit to the source counts, as
normalized via the detailed simulated source retrieval results of
\cite{Kim07b}. The {\it xskycover} table covers the full Extended
ChaMP with sky pixels, each $10' \times 10'$, whose boundaries are
chosen to match a regular commensurate grid across the sky.  The final
sensitivity in any given sky pixel is interpolated from the two
threshold maps computed at scales that bracket the size of the point
spread function (PSF) at that location \citep[for additional details,
see the Appendix to][]{Green09}. The accuracy of the {\it xskycover}
counts limits for 50\% and 90\% detection completeness were verified
by \cite{Aldcroft08} using the \Chandra\ Deep Field South (CDF-S).

\subsection{The Sloan Digital Sky Survey}
\label{sdss}

The Sloan Digital Sky Survey is an extensive photometric and
spectroscopic optical survey that covers nearly one-quarter of the
northern sky \citep{Stoughton02, Abazajian03, Abazajian04,
Abazajian05, Adelman-McCarthy06, Adelman-McCarthy07,
Adelman-McCarthy08}. With a dedicated 2.5m telescope the SDSS
accomplishes uniform photometry in five filters ($u,g,r,i,z$) to
magnitudes as faint as $r < 22.5$, with astrometric uncertainty
similar to \Chandra's \citep[\ie, better than 0.1\asec\ for $r <
20.5$; ][]{Pier03}. The SDSS data pipeline analyzes object morphology
and provides reliable star-galaxy separation to $r \sim 21.5$
\citep{Lupton02,Scranton02}. During its main survey phase (through
Data Release 7 [DR7]), the SDSS has obtained photometric measurements
for over 350 million unique objects. The SDSS also employs a pair of
dual multiobject fiber-fed spectrographs, and has obtained spectra for
\about 900,000 galaxies, \about 120,000 quasars, and \about 450,000
stars \footnote{See http://www.sdss.org for additional details and
public data access.}.

For the present study we query SDSS Data Release 5
\citep[DR5;][]{Adelman-McCarthy07}, \ie\ imaging and spectroscopy
taken through July 1, 2005, which includes photometry for 217 million
objects over 8000 $deg^2$ and about 1 million spectra of galaxies,
quasars, and stars chosen from 5713 $deg^2$ of the imaging data. By
selecting SDSS data within 20\amin\ of each \Chandra\ aimpoint (or
within 28\amin\ where the expansion yielded additional high-confidence
matches), we create a ChaMP/SDSS catalog of objects within the 323
\Chandra\ obsids described above. A photometric recalibration effort
for the SDSS Data Release 6 (DR6), resulted in improvements to the
SDSS photometric zero points \citep{Adelman-McCarthy08,Padman08}. We
thus cross-correlate our DR5 ChaMP/SDSS catalog with the DR6 to
utilize these ``UberCal'' magnitudes. As mentioned above, we also
cross-correlate our ChaMP/SDSS catalog with the new SDSS DR7
spectroscopic database to obtain the largest, most up-to-date
spectroscopic sample.

\section{Sample Selection} 
\label{sample_select}

\begin{figure*}[htb]
\includegraphics[width=0.9 \textwidth]{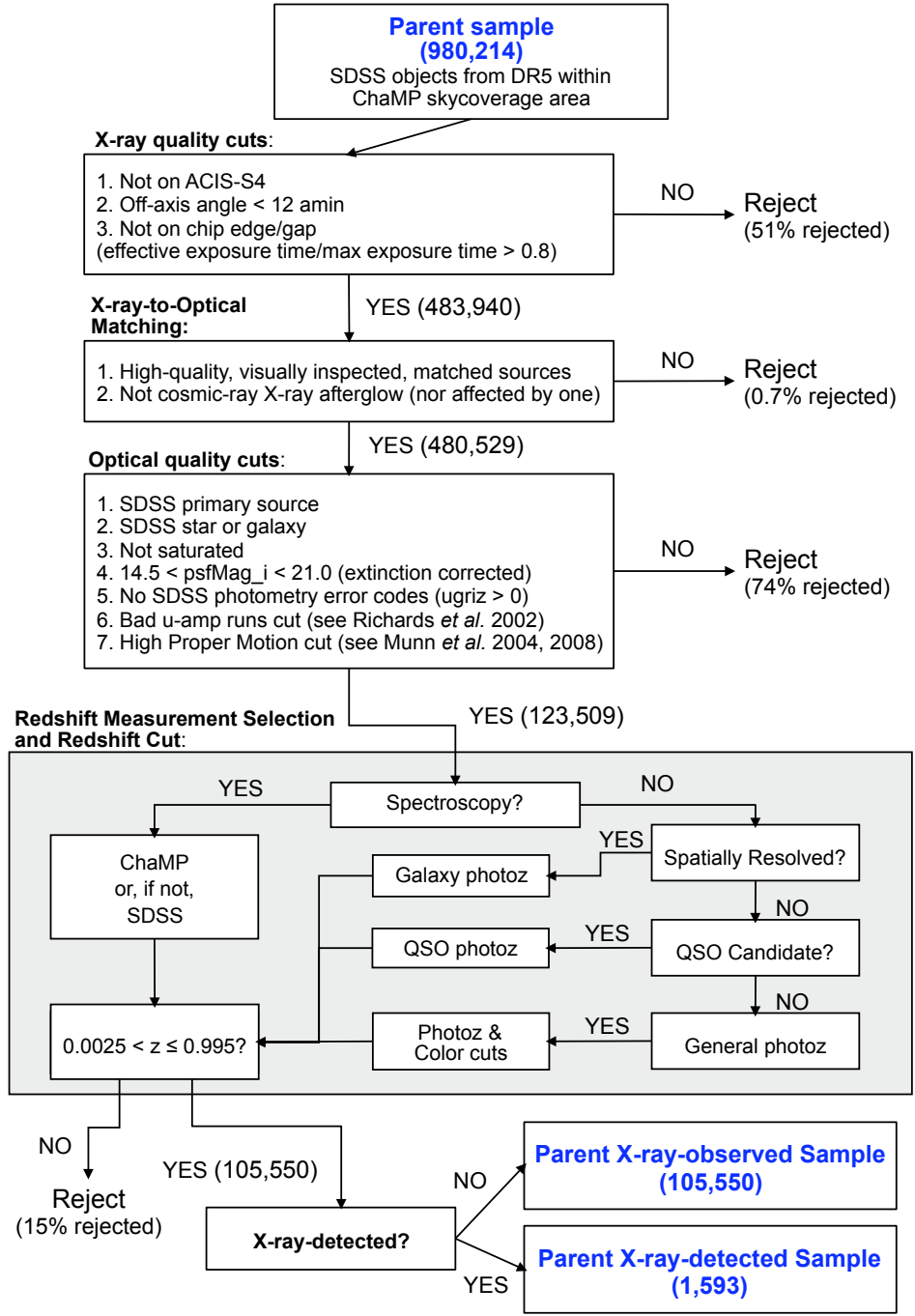}
\caption{Decision tree showing the sample selection criteria for clean
  samples of SDSS objects within the ChaMP sky coverage area for which
  \Chandra\ X-ray flux limits can be assigned, and for \Chandra\ X-ray
  detections. We refer to these as the ``Parent X-ray-observed'' and
  ``Parent X-ray-detected'' samples, respectively.}
\label{decision_tree}
\end{figure*}

To create a high-quality optical galaxy sample with an X-ray flux (or
flux limit) for each object, we apply a series of X-ray and optical
selection criteria to the ChaMP/SDSS sample. Our sample-selection
logic is summarized in Fig. \ref{decision_tree} and described in
detail below.

\subsection{X-ray and Optical Quality Cuts} 
\label{quality}

As described in \S \ref{champ}, the Extended ChaMP contains 392 unique
\Chandra\ fields, 323 of which fall within the SDSS DR5 footprint. The
catalog of SDSS objects recovered within this overlap region contains
980,214 individual SDSS objects; we refer to this as our ``Parent''
sample.

%--------------------------FIGURE 1---------------------------
%-------------------------------------------------------------

To insure the deepest and most uniform X-ray coverage, we exclude area
from the ACIS-S S4 chip ({\tt CCD 8}) which sustained damage early in
the \Chandra\ mission and has high background and streaking as a
result. We accept only area with off-axis angles (OAA) $< 12'$ --- at
larger off-axis angles the PSF broadens considerably and becomes
distorted \citep{Feigelson02, Kim07a}, which leads to large
uncertainties in centroiding and source counts.  We also exclude area
near ACIS chip edges and in chip gaps where the exposure times are
significantly shorter. To enforce this criteria, we require that the
effective exposure time at each {\it xskycover} sky pixel be greater
than 80\% of the maximum exposure time for the obsid. These X-ray
quality cuts reduce our X-ray sky coverage to \about $26~deg^2$.

In a number of the \Chandra\ fields the SDSS DR5 data does not cover
the full obsid. To crudely quantify the optical coverage of our X-ray
fields, we perform visual inspection of the 323 ChaMP/SDSS overlap
obsids and assign a coverage fraction to each. An estimate of the sky
area based on these rough coverage fractions is approximately
$21~deg^2$ and is a lower limit to our actual joint ChaMP-SDSS
coverage. The SDSS catalog associated with this sky coverage area
contains 483,940 objects (\about 49\% of the Parent sample).

For the X-ray-detected subset of the galaxy sample, we apply
additional quality constraints. All X-ray-detected objects in the
ChaMP survey area are visually inspected, wherein we overplot X-ray
centroids and their associated positions on both the X-ray and optical
images to identify poorly matched, multiply matched, or
photometrically contaminated objects. We accept only objects with the
highest-confidence, uncontaminated matches, \ie\ a single optical
counterpart with an X-ray-to-optical position offset no greater than
2\asec\ and/or less than the 95\% X-ray position error. Match
statistics for the X-ray detections in the Extended ChaMP catalog are
described in detail in \cite{Covey08} and \cite{Green09}. We also
eliminate X-ray sources that most likely result from (or have possibly
been affected by) cosmic-ray afterglows. The sources rejected by this
set of ChaMP quality cuts decreases our sample by less than 1\%.

An accurate AGN fraction requires a clean sample of both
X-ray-detections and non-detections. We thus also remove photometric
contaminants from our optical sample using the ``flags'' included in
the SDSS photometric catalogs\footnote{See
http://www.sdss.org/dr7/products/catalogs/flags.html for a detailed
description of the SDSS image processing flags.}. We include only
unsaturated, {\tt PRIMARY} (mode$=$1) objects that are classified as
either a star (type$=3$) or a galaxy (type$=6$) and whose 5-band
photometry is without error codes ($u,g,r,i,z > 0$). To avoid poor
photometry both at the bright and faint extremes of the survey, and to
insure uniform coverage at the faintest fluxes, we select objects with
$i$-band psf-magnitudes between 14.5 and 21.0. We also eliminate
objects impacted by bad $u$-amplifier runs \citep{Richards02}.

As described in \S \ref{sdss}, the SDSS pipeline provides reliable
star-galaxy separation for objects down to $r \sim 21.5$. This
distinction is a morphological one, and thus cannot correctly
characterize active galaxies and quasars that appear to be point
sources without consulting optical color information. Faint objects at
high redshift are also likely to be misclassified by morphological
diagnostics. Eliminating all SDSS ``stars'' would thus bias our
overall fraction. As a result, we do not remove point sources {\it a
priori} from our sample. Instead, we eliminate stars based on their
proper motions, positions in SDSS color-color and color-magnitude
diagrams, and spectroscopic or photometric redshifts. (See \S
\ref{redshift} below for a description of our color and redshift
cuts.)

\cite{Munn04,Munn08} have published a proper motion catalog that
combines USNO-B and SDSS observations. We use the Munn catalog to
assign proper motions to our catalog objects where there is a single
USNO-B match within $1''$ (match$=1$), the rms fit residuals (sigRa,
sigDec) are less than $350~mas$ in each coordinate, and there is at
least one USNO-B detection and one SDSS detection per source (nfit $>$
2). To eliminate high proper motion objects (\ie\ stars), we require
that the proper motion in at least one coordinate be larger than
$3~\sigma$, where $\sigma$ is the proper motion uncertainty in that
coordinate (\eg, $|PM_{RA}| \ge 3 \times \sigma_{RA}$), and that the
total proper motion be larger than 11~mas\,yr$^{-1}$.

The combined X-ray and optical quality cuts described in this section
result in a catalog of 123,509 SDSS objects.

\subsection{Redshift and Color Selection}
\label{redshift}

%--------------------------FIGURE 2---------------------------
\begin{figure}[bt]
\includegraphics[width=0.48 \textwidth]{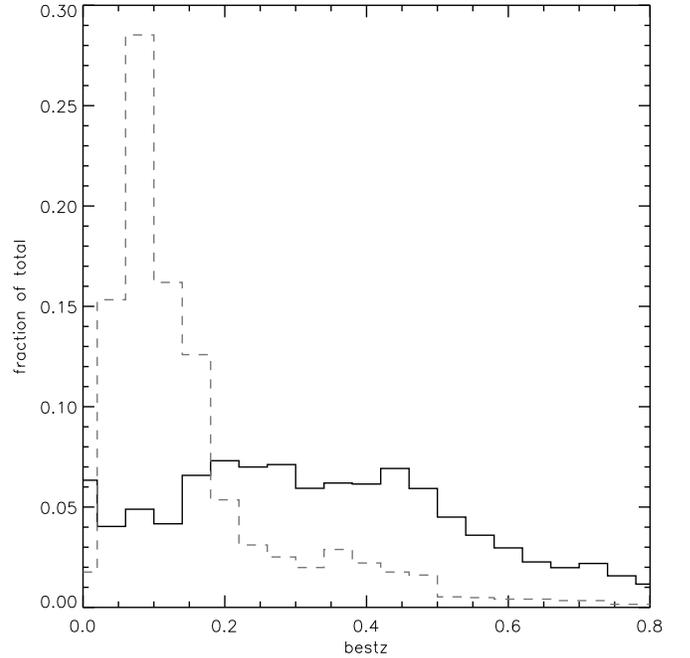}
\caption{Fractional histogram of the best redshift for the ``Full''
sample (105,550 galaxies; black solid line) and the spectroscopic
sub-sample (2668 objects from ChaMP and SDSS DR7 spectroscopy; gray
dashed line). The Full sample contains redshifts to $z \sim 1$,
where the photometric redshifts become less reliable, while the
spectroscopic sub-sample shows a strong peak at low-redshift, $0.05 < bestz
< 0.1$. The peak in the spectroscopic sub-sample results largely from
SDSS spectroscopic selection criteria.}
\label{bestz_hist}
\end{figure}
%-------------------------------------------------------------

To assign a redshift to each object in our Parent sample we exploit
both spectroscopic and photometric redshift catalogs. We select a
\bestz, the highest quality redshift available, by ranking the
redshift catalogs as follows: (1) the ChaMP spectroscopic catalog
\citep[398 objects;][]{Green04}; (2) the SDSS DR7 spectroscopic
catalog \citep[2,270 objects;][]{Abazajian09}; (3) the SDSS neural
network (NN) \photoz2\ catalog for resolved objects \citep[68,908
objects;][]{Oyaizu08}; (4) the DR6 Quasar Photometric Redshift Catalog
\citep[281 objects;][]{Richards09}; and (5) for unresolved point
sources with no other redshift determination, the DR6 version of the
SDSS {\it Photoz} catalog, derived from the template-fitting method
with repaired interpolated templates \citep[33,693
objects;][]{Csabai03}. Properties of the two SDSS photometric redshift
catalogs, {\it Photoz} and \photoz2, are summarized in Table
\ref{photoz_tab}. We choose the DR6 version of the SDSS {\it Photoz}
catalog (henceforth, the ``point source template'' or ``template''
photo-$z$) because it contains photometric redshift estimates for
point sources that are not included in the other catalogs; \eg, the
newer DR7 catalog is limited to objects classified morphologically by
the SDSS pipeline as galaxies \citep{Abazajian09}. Based on our
selection scheme, only objects with stellar morphology are assigned
point source template photo-$z$'s. We eliminate all
spectroscopically-confirmed stars from our sample.

\begin{figure}[tb]
\includegraphics[width=0.48 \textwidth]{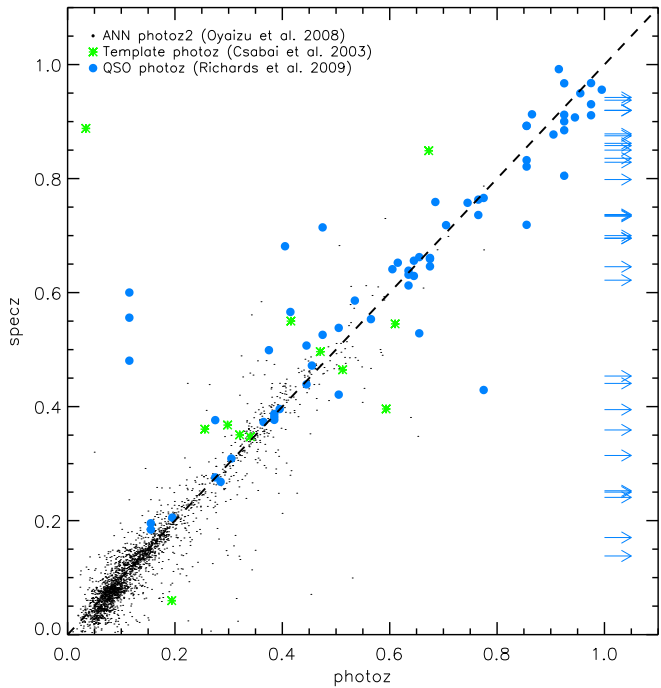}
\caption{Spectroscopic redshift versus photometric redshift for 2668
objects in our galaxy sample with spectroscopy (398 from ChaMP and
2270 from SDSS DR7). The symbol type represents the photometric
redshift catalog used: the SDSS DR6 resolved NN photo-$z$ galaxy
catalog \citep[][black points]{Oyaizu08}, the SDSS DR6 QSO catalog
\citep[][blue filled circles and blue arrows]{Richards09}, the SDSS
DR6 point source template photo-$z$ catalog \citep[][green
stars]{Csabai03}. The most prominent photo-$z$ failures at redshifts
above $z \sim 1$ are for QSOs (see \S \ref{mc}). \\ 
(A color version of this figure is available in the online journal.)}
\label{photz_specz}
\end{figure}

Figure \ref{bestz_hist} shows normalized histograms of the redshifts
for our ``Full'' sample (solid black line) and for the spectroscopic
sub-sample (gray dashed line). For those objects with both photometric
and spectroscopic redshifts, 97\% have ${\Delta~z \over (1+z)} < 0.1$
(Fig. \ref{photz_specz}). The photo-$z$ algorithms are the least
reliable for quasars (Fig. \ref{photz_specz}, filled circles and
arrows). We discuss the impact of these potentially large photometric
redshift errors on our determination of the AGN fraction in \S
\ref{mc}.

%--------------------------FIGURE 3---------------------------
%-------------------------------------------------------------

To quote a single error estimate on our \bestz\ value, we utilize
and/or convert the redshift errors in each catalog into a $1~\sigma$
Gaussian error.  ChaMP spectroscopic redshifts and $1~\sigma$ redshift
errors are estimated via radial velocities, using the
cross-correlation and emission line fitting techniques in the IRAF
$rvsao$ task \citep{Kurtz98,Green04}. Redshifts and their errors for
the SDSS DR7 spectroscopic sample are estimated using emission-line or
cross-correlation techniques\footnote{A discussion of SDSS redshift
determinations is available at
http://www.sdss.org/dr7/algorithms/redshift\_type.html.}.

The \cite{Richards09} QSO photometric redshift catalog uses a Bayesian
approach to estimate a most likely redshift ($zphot$), as well as a
high and low redshift ($zphothi$ and $zphotlo$) which represent an
approximate range for $zphot$, and a probability that $zphot$ is
within this range. This photometric redshift determination is
described in detail by \cite{Weinstein04}. We estimate $1~\sigma$
errors (\ie\ 68\% probability intervals) by approximating the $zphot$
errors as symmetric and Gaussian. We assume that the $zphothi$ and
$zphotlo$ values bracket the $zphotprob$ range for a Gaussian
distribution and derive a ``$1~\sigma$'' error using Gauss' formula
(\ie\ using the error function, {\it erf}). We also utilize the full
redshift probability distribution function for the QSO subset in the
Monte Carlo simulations described in \S \ref{mc}.

%---------------------------TABLE 1---------------------------
% -------------------------------- PHOTOZ TABLE ----------------------------------

\begin{deluxetable}{lcc} 
\tabletypesize{\scriptsize}
\tablewidth{0pt}
\tablecaption{SDSS Photometric Redshifts}
\tablehead{\colhead{Catalog} & \colhead{Photoz} & \colhead{Photoz2}}

\startdata
%Catalog       &  Photoz           &  Photoz2 \\
Number        &  33,693		  &  68,908 \\
Source type   &  unresolved	  &  resolved \\
Method        &  template fitting &  neural network \\
Abbreviation  &  template         &  NN \\
Reference     &  \cite{Csabai03}  &  \cite{Oyaizu08}
\enddata

\label{photoz_tab}
\vspace{2mm}
\end{deluxetable}

%\label{photoz_tab}
%-------------------------------------------------------------

Both the SDSS resolved NN photo-$z$ and point source template photo-$z$
catalogs include Gaussian $1~\sigma$ errors, though the methods used
to derive them differ. In the resolved NN photo-$z$ analysis,
\cite{Oyaizu08} use a nearest neighbors estimator, with a
spectroscopic training set, to determine their best photometric
redshift and the associated error. We select their ``CC2'' values,
which employ both magnitudes and concentration indices in the redshift
determination. The point source template photo-$z$ catalog uses a
template-fitting algorithm with repaired interpolated templates to
determine the photometric redshift and associated 68\% uncertainty,
assuming Gaussian errors \citep{Csabai03}.

%--------------------------FIGURE 4---------------------------
\begin{figure*}[htb]
\includegraphics[width=1.0 \textwidth]{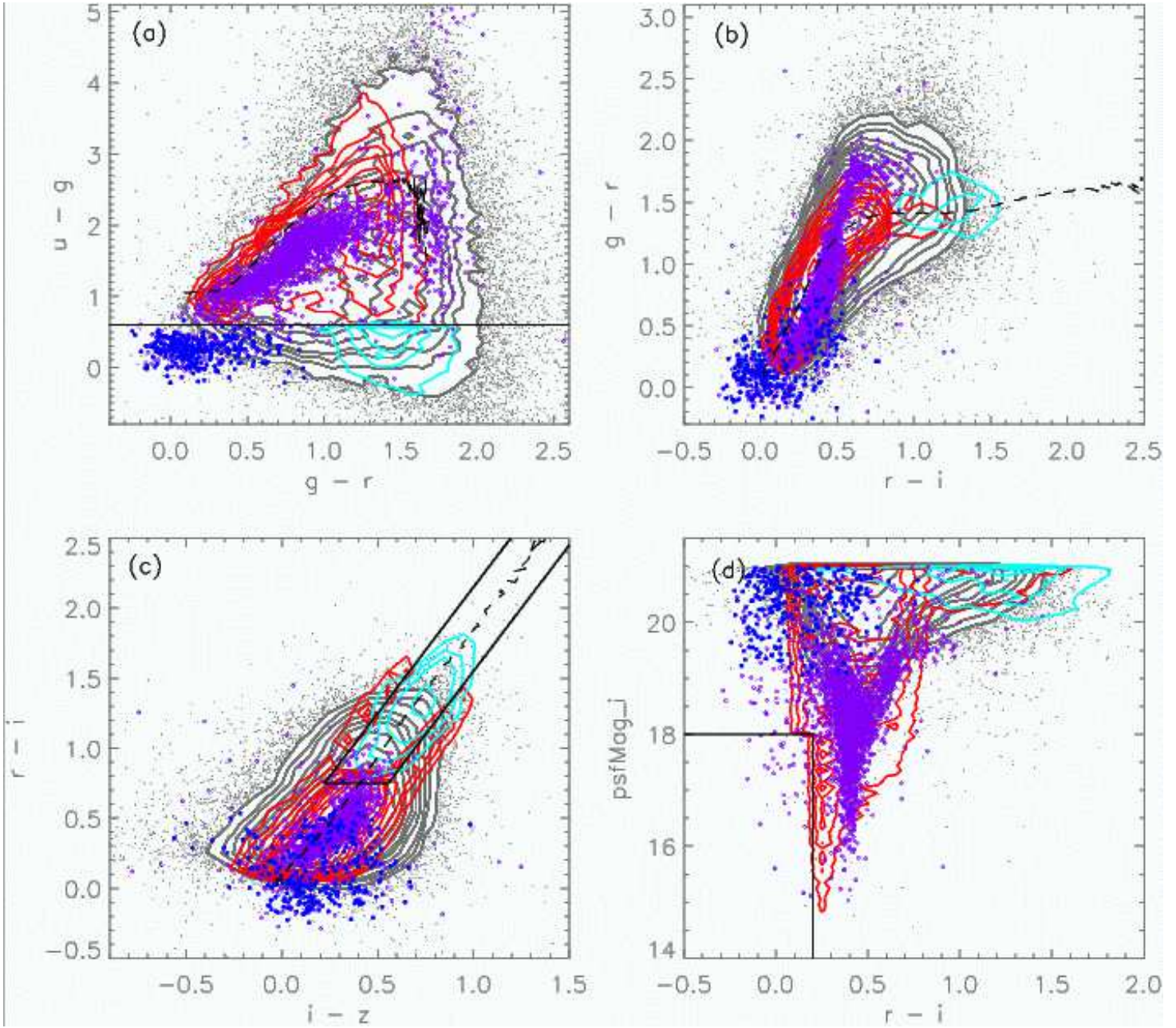}
\caption{Color-color and color-magnitude diagrams for the Full sample
after the cuts described in \S \ref{redshift}, and outlined in
Fig. \ref{decision_tree}, have been applied to remove galactic
(stellar) contaminants, without eliminating extragalactic
objects. Gray contours/dots correspond to objects assigned resolved NN
photo-$z$'s from the \cite{Oyaizu08} catalog. Red and cyan contours
represent objects that have point source template photo-$z$'s
\citep{Csabai03} and $(u-g)$ colors above and below 0.6 (panel a,
solid black line), respectively. Dark blue stars indicate objects with
QSO photo-$z$'s from \cite{Richards09}. Objects with spectroscopy are
indicated with open purple circles. The solid black lines in panels
c and d indicate regions from which red ($(u-g)>0.6$) point sources
with template photo-$z$'s have been eliminated. The black dashed lines
in panels a-c show the SDSS stellar locus from \cite{Covey07}.}
\label{color_color}
\end{figure*}
%-------------------------------------------------------------

For the sample that has been assigned a point source template
photo-$z$, we make additional color and magnitude cuts to reduce our
contamination from stars.  We accept all objects for which $(u-g) \le
0.6$ (Fig. \ref{color_color}a, cyan contours), as these are likely to
be extragalactic objects associated with a star forming or otherwise
active galaxies \citep[see, for example, ][]{Richards02}. For objects
with $(u-g) > 0.6$, we remove objects from two regions in color-color
and color-magnitude space that are populated primarily by stars (solid
black lines, Figs. \ref{color_color}c and \ref{color_color}d): (1) the
first elimination region is defined by SDSS psfMag\_i $\le 18$ and
$(r-i) \le 0.2$; (2) the second is an area along the main sequence
stellar locus in the $(r-i)$ vs. $(i-z)$ diagram \cite[parametrized
by][black dashed line, Fig. \ref{color_color}c]{Covey07}. For the
second cut, we approximate the main sequence stellar locus as a
straight line, described by the formula $(r-i) = 1.8 \times (i-z) +
0.05$. We remove objects with $(r-i) > 0.75$ in a diagonal region
within $\pm~0.3$ of the stellar locus center line (solid black lines,
Fig. \ref{color_color}c). We do not cut blueward of $(r-i) = 0.75$,
because our spectroscopic sample (open purple circles), which consists
of galaxies and AGN only, indicates that this would begin to remove
galaxies from our sample. Despite our efforts to eliminate stellar
point sources, our sample is still likely to suffer from considerable
stellar contamination. We discuss and quantify the impact of this
contamination on our AGN fraction in \S \ref{bias}.

Once a \bestz\ has been assigned and the most likely stellar
contaminants have been removed, we restrict our attention to those
objects with $0.0025 < bestz < 0.995$. This redshift cut avoids cases
where the photometric errors are large and/or the photometric redshift
algorithms cannot find a suitable fit.  These redshift, color, and
spectroscopic cuts remove another \about 15\% of the sample, leaving
105,550 objects in our optical galaxy catalog --- the ``Parent
X-ray-observed'' sample. Of these, 1,593 are X-ray detected --- these
form our ``Parent X-ray-detected'' sample.  (As a result of this
selection scheme, the objects in the X-ray-detected sample are a
subset of those in the X-ray-observed sample.) These objects are shown
in the color-color and color-magnitude diagrams of
Fig. \ref{color_color}: ChaMP/SDSS spectroscopic redshifts (open
purple circles); QSO photometric redshifts (dark blue stars); resolved
NN photo-$z$ galaxies (gray contours); the point source template
photo-$z$ population with $(u-g) \le 0.6$ (cyan contours); and the
point source template photo-$z$ population with $(u-g) > 0.6$, which
also meet the color cuts described above (red contours).

\vspace{5mm}

\section{Derived Properties and Volume-Limited Samples}
\label{properties}

\subsection{Galaxy Properties}
\label{derived}

We calculate optical k-corrections and absolute magnitudes using {\tt
kcorrect v4\_1\_4} \citep{Blanton07}, via the relation
\begin{equation}
m_R = M_Q + DM(z) + K_{QR}(z) - 5~{\rm log}~h, 
\end{equation}

\begin{figure}[tb]
\includegraphics[width=0.5 \textwidth]{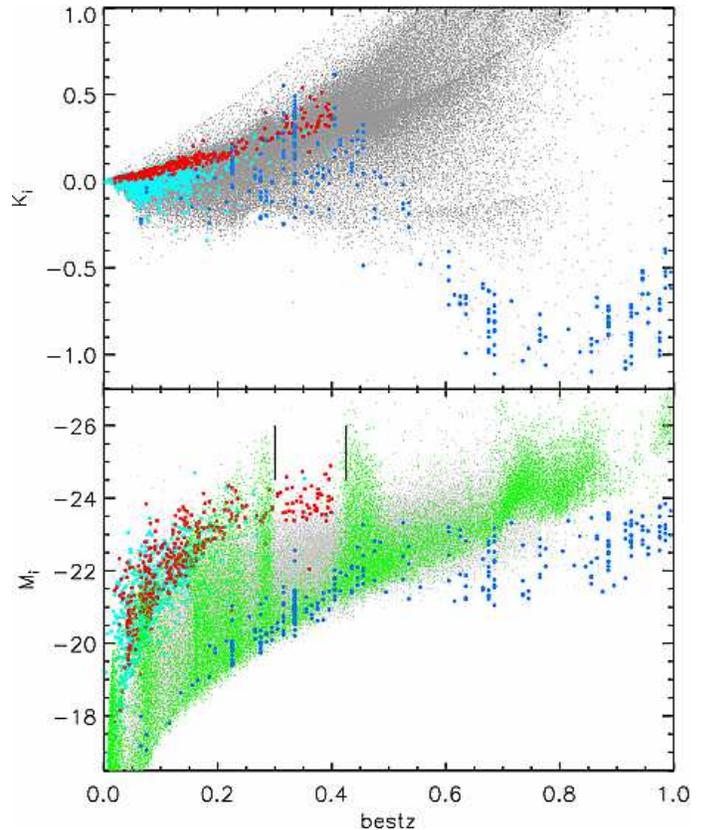}
\caption{ ({\it Top}) K-correction as a function of \bestz\ for the
Full galaxy sample (gray points). Objects with spectroscopic
classifications from the DR7 MPA/JHU catalog are marked (red stars for
absorption line galaxies, cyan stars for narrow emission line
galaxies), and objects with QSO photometric redshifts (blue filled
circles) are also overplotted for comparison.  ({\it Bottom}) Absolute
magnitude as a function of \bestz, determined after the k-correction
from the top panel has been applied. The light gray points indicate
objects assigned a resolved NN photo-$z$ and the green points are
those with a point source template photo-$z$ (\ie, the sum of the red
and cyan contours in Fig. \ref{color_color}); the remaining symbols
are as in the top panel. The prominent structures at \bestz\ \about
0.3 and 0.43 result from the 4000 \AA\ Ca H-K break passing between
the SDSS $g$ and $r$ filters \citep[see, \eg,][]{Padman07}. We
estimate the edges of this filter gap at 5200 \AA\ and 5700 \AA\ and
mark the redshift corresponding to observed-frame 4000 \AA\ as black
vertical lines.}
\label{kcorr}
\end{figure}

\noindent{where $m_R$ is the apparent magnitude, $M_Q$ is the absolute
magnitude, $DM(z)= 25 + 5~{\rm log}~[D_L/(h^{-1}$ Mpc)] is the
distance modulus calculated from the luminosity distance $D_L$, and
$K_{QR}(z)$ is the k-correction \citep{Hogg02}. In particular, we use
{\tt sdss\_kcorrect} and input our \bestz, extinction-corrected SDSS
model-magnitudes, and model-magnitude errors. Since the {\tt kcorrect}
algorithm assumes a Hubble constant $H_0 =
100~h$~km\,s$^{-1}$\,Mpc$^{-1}$ with $h = 1$, we also convert to $h =
0.7$. The resulting $i$-band k-corrections (to rest-frame $z=0$) and
absolute magnitudes, $M_i$, are shown as a function of \bestz\ in
Fig. \ref{kcorr}.  We identify spectroscopic absorption line galaxies
(red stars) and narrow emission line galaxies (cyan stars) based on
their MPA/JHU line classifications. The spectroscopic sub-sample
appears at the bright end of the absolute magnitude distribution, as
expected from SDSS main galaxy sample target selection algorithm which
requires an $r$-band Petrosian magnitude brighter than 17.77. In the
top panel of Fig. \ref{kcorr} the absorption line galaxies follow a
tighter trend in k-correction than their emission line
counterparts. This trend is also expected, since emission line
strengths vary widely in star-forming galaxies and AGN. We indicate
objects with QSO photometric redshifts (blue filled circles) and note
that in many cases their k-corrections (and resulting $M_i$) lie far
from the main galaxy locus. We test the impact of this bias on our
fractions in \S \ref{mc} and discuss its source in \S \ref{bias}.}

%--------------------------FIGURE 5---------------------------
%-------------------------------------------------------------

In the bottom panel of Fig. \ref{kcorr} we color-code the points by
the redshift catalog from which they originate (red and cyan symbols
as in the top panel for spectroscopic redshifts, blue filled circles
for QSO photo-$z$'s, light gray points for the resolved NN photo-$z$
catalog, and green points for the point source template photo-$z$
catalog). There are several noticeable structures at the bright end of
the $M_i$ vs. \bestz\ relation. The most prominent of these results
from the Ca H-K 4000 \AA\ break passing between the SDSS $g$ and $r$
filters \citep[see, for example, Fig. 1 of][]{Padman07}. We estimate
the edges of this ``gap'' at 5200 \AA\ and 5700 \AA\ and mark the
redshift corresponding to observed-frame 4000 \AA\ as black vertical
bars. In this rendering of the data, we see that most of the structure
comes from the point source template photo-$z$ objects, which are, by
definition, red point sources ($(u-g) > 0.6$) in the SDSS. That the
point source template photo-$z$ catalog does not perform well for
these objects is not a surprise since its primary spectroscopic
training sets are the SDSS main galaxy sample and the luminous red
galaxy sample.

%--------------------------FIGURE 6--------------------------
\begin{figure*}[htb]
\includegraphics[width=1.0 \textwidth]{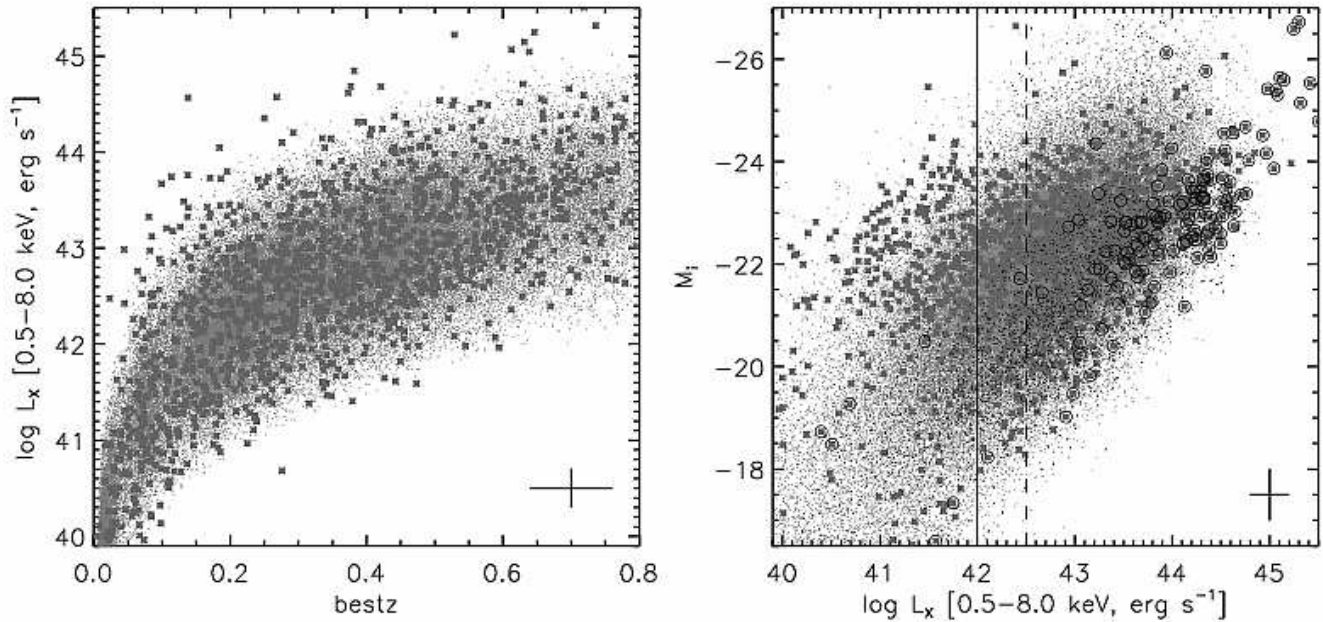}
\caption{({\it Left}) Log of the broad-band (0.5-8.0~keV) X-ray
  luminosity, log $L_X$, as a function of \bestz\ for the
  X-ray-observed (light gray points) and X-ray-detected samples (dark
  gray stars). ({\it Right}) Absolute $i$-band magnitude vs. log $L_X$
  (gray symbols as in the left panel). The solid line marks log $L_X =
  42$, the limiting X-ray luminosity used for our AGN fraction
  calculations; the dashed line indicates log $L_X = 42.5$, the limit
  used for AGN fraction calculations in the zCOSMOS survey
  \citep{Silverman09a}. The black points (upper limits) and large,
  black open circles (detections) mark objects that reside off the
  main locus in the $M_i$ vs. host mass diagram (see
  Fig. \ref{mass_mi} and \S \ref{bias}). Representative $1~\sigma$
  error bars are plotted in the lower right corner of both
  panels. These are the median uncertainties for the X-ray-observed
  sample, derived from the 100-iteration ``full'' Monte Carlo
  simulation described in \S \ref{mc} --- the error bars thus derive
  only from uncertainties in the redshifts.}
\label{mi_lx_z}
\end{figure*}
%-------------------------------------------------------------

Using the ChaMP's X-ray sensitivity maps (see description of the {\it
xskycover} table in \S \ref{champ}), we assign a \Chandra\ broad-band
(0.5-8.0~keV) X-ray flux limit to each of the galaxies in our
X-ray-observed sample. We also assign a broad-band X-ray flux to each
detection. With these X-ray fluxes (and flux limits), we calculate the
X-ray-to-optical flux ratio, {\it \`a la} \cite{Horn00}, using the SDSS
$r$-band apparent model-magnitudes:
\begin{equation}
{\rm log}~(f_x/f_{r'})= {\rm log}~f_x(0.5-2 {\rm keV}) + 5.57 + (r'/2.5)~,
\end{equation}

\noindent{where the conversion constant is determined assuming
$f(\nu_0) = 3.631 \times 10^{-20}$~erg\,cm$^{-2}$\,s$^{-1}$\,Hz$^{-1}$
and using the SDSS $r'$-band effective wavelengths \citep{Fukugita96}.
We also use the \bestz, together with ChaMP X-ray fluxes and flux
limits, to calculate luminosity distances\footnote{Luminosity and
comoving distances are calculated using the Python version of the
Wright Cosmology Calculator \citep{Wright06}.} ($D_L$), broad-band
X-ray luminosities ($L_X$) for detections, and X-ray luminosity limits
\citep[$L_{X,lim}$; 90\% confidence in the broad band,][]{Aldcroft08}
for non-detections. In Fig. \ref{mi_lx_z} we show log $L_X$ (dark gray
stars) and log $L_{X,lim}$ (light gray points) in units of
erg\,s$^{-1}$. On the left, the X-ray luminosities (limits) are shown
as a function of \bestz\ --- the X-ray-detections span the locus of
optical galaxies uniformly, indicating that our optical selection has
not introduced obvious selection biases in the X-ray. On the right, we
show $M_i$ vs. log $L_X$ (or log $L_{X,lim}$) and mark two luminosity
limits, log $L_X = 42$ and log $L_X = 42.5$, which we employ later in
calculating our field AGN fraction and comparing our results to those
of deep-field surveys.}

\begin{figure}[ht]
\includegraphics[width=0.5 \textwidth]{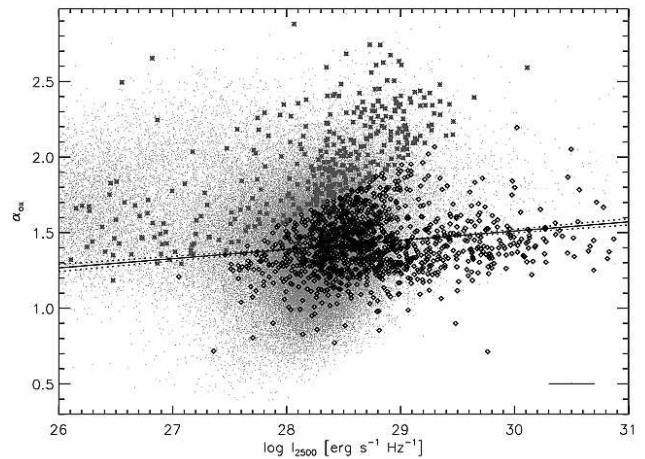}
\caption{$\alpha_{ox}$ vs. optical (2500 \AA) luminosity for the Full
galaxy sample. X-ray-observed galaxies are shown as light gray points,
X-ray-detections are marked with large symbols according to their
broad-band X-ray-luminosity: log $L_X \ge 42$ (black diamonds) and log
$L_X < 42$ (dark gray stars). The ChaMP QSO best-fit regression line
($\alpha_{ox} = (0.061 \pm 0.009) {\rm log}(l_{2500 {\rm \AA}}) -
(0.319 \pm 0.258)$) from \cite{Green09} is shown as a black solid line
with errors. Representative $1~\sigma$ error bars are as described in
Fig. \ref{mi_lx_z}; the median error bar for the $\alpha_{ox}$ is
smaller than the width of the horizontal line for the ${\rm
log}(l_{2500 {\rm \AA}})$ error bar.}
\label{olum_aox}
\end{figure}

Assuming an optical continuum power-law slope of $\alpha = -0.5$
($f_{\nu} \propto \nu^{\alpha}$, where $\nu$ is the emission
frequency), we derive rest-frame monochromatic optical luminosities at
2500 \AA\ ($l_{2500 {\rm \AA}}$;~erg\,s$^{-1}$\,Hz$^{-1}$) using the
SDSS $g$-band dereddened magnitude. In theory the $u$-band magnitude
would be more appropriate for objects with $bestz < 0.53$, because the
filter's central wavelength is closer to $(1 + z) \times 2500 {\rm
\AA}$, but the large errors on $u$-band magnitudes mitigate this
improvement. We derive the 2~keV luminosity ($l_{2 {\rm
keV}}$;~erg\,s$^{-1}$\,Hz$^{-1}$) from the broad-band flux using PIMMS
and assuming an unabsorbed power-law with spectral slope
$\Gamma=1.7$.  The slope of a hypothetical power law from 2500 \AA\ to
2~keV is commonly characterized by $\alpha_{ox}$, where $\alpha_{ox} =
0.3838 \times {\rm log}~(l_{2500 {\rm \AA}}/l_{2 {\rm keV}})$; we plot
$\alpha_{ox}$ as a function of $l_{2500 {\rm \AA}}$ in
Fig. \ref{olum_aox}.

%--------------------------FIGURE 7---------------------------
%-------------------------------------------------------------

\subsection{Volume-Limited Samples}
\label{vcomplete}

\begin{figure}[t]
\includegraphics[width=0.5 \textwidth]{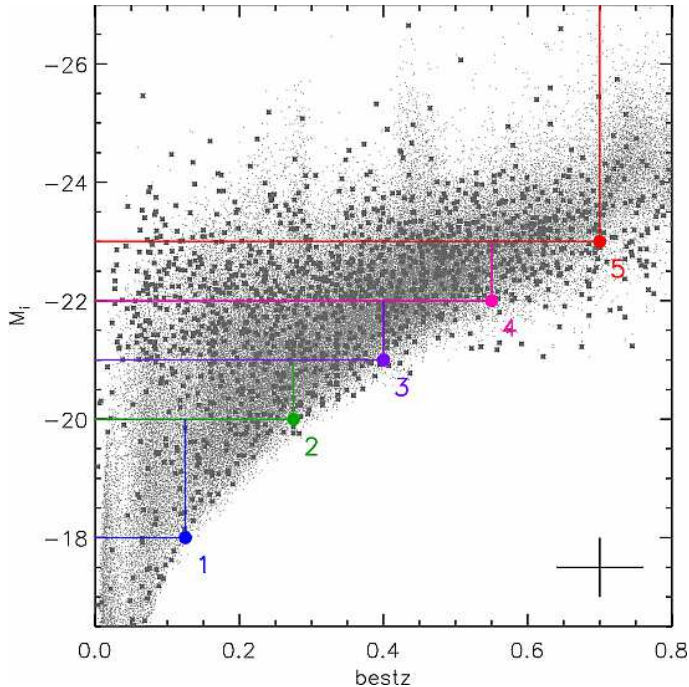}
\caption{Absolute $i$-band magnitude vs. \bestz\ for the 105,550
galaxies in the X-ray-observed sample (light gray points) and the
1,593 X-ray-detected sources (large dark gray stars). Colored lines
and large symbols/numbers indicate the five independent,
volume-limited samples (see \S \ref{vcomplete} and Table
\ref{samples}) for which we determine the X-ray-active and AGN
fractions. Representative $1~\sigma$ error bars are as described in
Fig. \ref{mi_lx_z}. \\
(A color version of this figure is available in the online journal.)
}
\label{Mi_z}
\end{figure}

Using these derived properties, we define five samples complete in
\bestz\ and $M_i$ (referred to as volume-limited samples
hereafter) to calculate our X-ray-active and AGN fractions. We note
that this makes ours an optically-selected, X-ray-detected (or
limited) study. The five volume-limited samples shown in
Fig. \ref{Mi_z} are defined as: (1) $z \le 0.125$ and $-18 \ge M_i >
-20$; (2) $z \le 0.275$ and $-20 \ge M_i > -21$; (3) $z \le 0.4$ and
$-21 \ge M_i > -22$; (4) $z \le 0.55$ and $-22 \ge M_i > -23$); and
(5) $z \le 0.7$ and $M_i \le -23$. (The low-redshift limit for all
samples is $z = 0.0025$.)  Their constituents are detailed in Table
\ref{samples}.

We determine each galaxy's stellar mass via the {\tt kcorrect} tool,
which generates the coefficients of a fit between a galaxy's spectral
energy distribution and a set of template spectra. \cite{Blanton07}
derive five global templates from 450 instantaneous bursts of star
formation from \cite{Bruzual03} spectral synthesis models, using the
\cite{Chabrier03} stellar initial mass function and the Padova 1994
isochrones, and 35 templates from MAPPINGS-III \citep{Kewley01} models
of emission from ionized gas.  These global templates are
characterized using ``nonnegative matrix factorization,'' which
determines a set of nonnegative templates that can be combined
linearly to explain a data set. The five global templates correspond
roughly to a very old galaxy template spectrum, a very young template,
and several intermediate templates \citep[see][Figure
4]{Blanton07}. The {\tt kcorrect} template fits can be interpreted
physically and we utilize them to determine mass estimates for each of
our galaxies. We sum the fit coefficients (units:
$1~M_{\odot}/(D_L/10~{\rm pc})^2$, where $D_L$ is the luminosity
distance) and multiply by $(D_L/10~{\rm pc})^2$ to find the total
current stellar mass for each galaxy. In the left panel of
Fig. \ref{mass_z}, we plot log $M_{\star}$, in units of solar mass, as
a function of \bestz; on the right we show rest-frame $(u-r)_0$
vs. log $M_{\star}$. To compare with studies that use mass to define
their galaxy samples \citep[\eg,
][]{Silverman08a,Silverman09a,Silverman09b}, we create complete mass
samples, denoted samples 1* -- 5*, analogous to those in
Fig. \ref{Mi_z}: $9 \le {\rm log}~M_{\star} < 9.8$, $9.8 \le {\rm
log}~M_{\star} < 10.4$, $10.4 \le {\rm log}~M_{\star} < 11$, $11 \le
{\rm log}~M_{\star} < 11.6$, and ${\rm log}~M_{\star} \ge 11.6$ (the
redshift intervals are unchanged). These mass intervals are selected
based on a crude correspondence between $M_i$ and log $M_{\star}$,
\ie\ one magnitude in $M_i$ is roughly equivalent to 0.4 dex in log
$M_{\star}$ (see also Fig. \ref{mass_mi}).

%---------------------------TABLE 2---------------------------
%-------------------------------- SAMPLE DESCRIPTION TABLE ----------------------------------

\begin{deluxetable}{lccccccccc}
\tabletypesize{\scriptsize}
\tablewidth{0pt}
\tablecaption{Volume-Limited Samples}
\tablehead{\colhead{} & 
\colhead{$bestz$} &
\multicolumn{2}{c}{$M_i$} & &
\multicolumn{2}{c}{Full} & &
\multicolumn{2}{c}{Spec} \\
\cline{3-4}\cline{6-7}\cline{9-10}
\colhead{S} & 
\colhead{max} &
\colhead{min} & 
\colhead{max} & &
\colhead{N$_{xdet}$} & 
\colhead{N$_{xlim}$} & &
\colhead{N$_{xdet}$} & 
\colhead{N$_{xlim}$}
}

\startdata
 1 & 0.125 & -18 &  -20 &&  29 &  5028 &&  2 & 182  \\
 2 & 0.275 & -20 &  -21 &&  88 & 10949 &&  7 & 342  \\
 3 & 0.400 & -21 &  -22 && 239 & 17984 && 37 & 566  \\
 4 & 0.550 & -22 &  -23 && 416 & 22095 && 68 & 567  \\
 5 & 0.700 & -23 &\ldots&& 323 & 13628 && 57 & 464  \\ 
\enddata

\tablecomments{Properties of the five volume-limited samples for the
Full galaxy catalog (described in \S \ref{vcomplete} and shown in
Fig. \ref{Mi_z}) and the Spectroscopic sub-sample.  The first column
gives the sample number; the second gives the maximum redshift
($bestz$; all samples have a minimum $bestz = 0.0025$); the third and
fourth give the absolute SDSS $i$-band magnitude range; and the
remaining columns contain the number of X-ray-detected and
X-ray-observed, optical galaxies.}
\label{samples}
\end{deluxetable}

%---------------------------------------------------------------------------

%\label{samples}
%-------------------------------------------------------------

\section{The Fraction}
\label{the_frac}

\subsection{The X-ray-Active Fraction}
\label{xactive_frac}

We calculate the fraction of X-ray-active galaxies for each of the
five samples complete in \bestz\ and $M_i$ described in \S
\ref{vcomplete}. In addition to the optical criteria, we apply X-ray
selection criteria so that our samples are complete in both the
optical and in X-rays. For each fraction, we define an X-ray
threshold, $L_X^{\prime}$, which gives the brightest X-ray luminosity
to which all objects in the sample are, or would have been, detected
by \Chandra. Any \emph{detected} source in the sample must have an
X-ray luminosity at or above the X-ray luminosity limit (at the 90\%
confidence level).

The details of the X-ray criteria are: (1) a galaxy appears in the
denominator only if its broad-band X-ray luminosity upper limit
$L_{X,lim}$ (see \S \ref{derived} for a definition) is less than the
chosen threshold, $L_X^{\prime}$, regardless of X-ray-detection
status. In other words, each optical galaxy counted in the denominator
would have been detected at $L_X^{\prime}$ since
$L_X^{\prime}>L_{X,lim}$, where $L_{X,lim}=4\pi\,D_L^2\,f_{X,lim}$.
(2) The numerator is the X-ray-detected subset of the denominator,
where the broad-band X-ray luminosity, $L_X$, for each detection is
greater than or equal to the same limiting $L_X^{\prime}$ used to
define the denominator. (We discuss the objects that appear in the
volume-limited optical samples, but do not pass the X-ray cut in the
Appendix.)

We can express these criteria mathematically with two functions, (1)
$G(L_{X,lim}, M_i, bestz)$ to describe galaxies with X-ray luminosity
limits, and (2) $X(L_X, L_{X,lim}, M_i, bestz)$ to describe the X-ray
detections. We write the denominator and the numerator of the fraction
as:
\begin{equation}
N_{lim} = {\sum^{z_{max}}_{z_{min}} \sum^{M_{bright}}_{M_{faint}} \sum^{L_X^{\prime}}_0 G(L_{X,lim},M_i,bestz)} ~,
\end{equation}

\begin{equation}
N_{det} = {\sum^{z_{max}}_{z_{min}} \sum^{M_{bright}}_{M_{faint}} \sum^{L_X^{\prime}}_0 \sum^{\infty}_{L_X^{\prime}} X(L_X,L_{X,lim},M_i,bestz)} ~,
\end{equation}

\noindent{where $z_{min}$, $z_{max}$, $M_{faint}$, and $M_{bright}$,
define our volume-limited optical samples, and $L_X^{\prime}$
establishes X-ray completeness. All five volume-limited samples have a
minimum redshift $z_{min} = 0.0025$.}

%--------------------------FIGURE 8---------------------------
%-------------------------------------------------------------

%--------------------------FIGURE 9--------------------------
\begin{figure*}[htb]
\includegraphics[width=1.0 \textwidth]{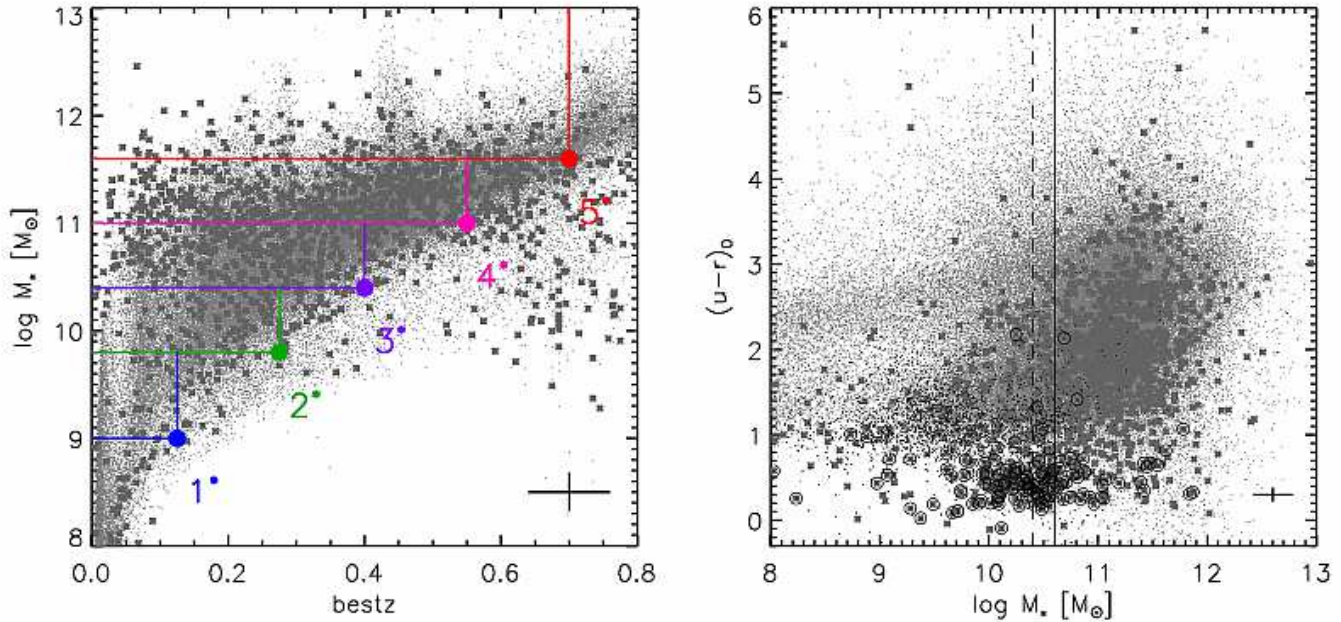}
\caption{({\it Left}) Log of the host galaxy mass, log $M_{\star}$,
  vs. \bestz\ (gray symbols as in Fig. \ref{Mi_z}). A set of samples,
  complete in galaxy mass, are indicated with colored lines and large
  symbols/numbers (see \S \ref{vcomplete}). These are analogous to the
  volume-limited samples in Fig. \ref{Mi_z}. ({\it Right}) Rest-frame
  $(u-r)_0$ color vs. log $M_{\star}$ (gray and black symbols as in
  Fig. \ref{mi_lx_z}). Black lines mark log $M_{\star} = 10.6$ (solid)
  and $10.4$ (dashed) and are used to define mass limits for a
  comparison between our field fractions and those of other X-ray
  surveys (see \S \ref{discussion}). Representative $1~\sigma$ error
  bars in both panels are as described in Fig. \ref{mi_lx_z}. \\
  (A color version of this figure is available in the online journal.)}
\label{mass_z} 
\end{figure*}
%-------------------------------------------------------------

%--------------------------FIGURE 10---------------------------
\begin{figure*}[htb]
\includegraphics[width=1.0 \textwidth]{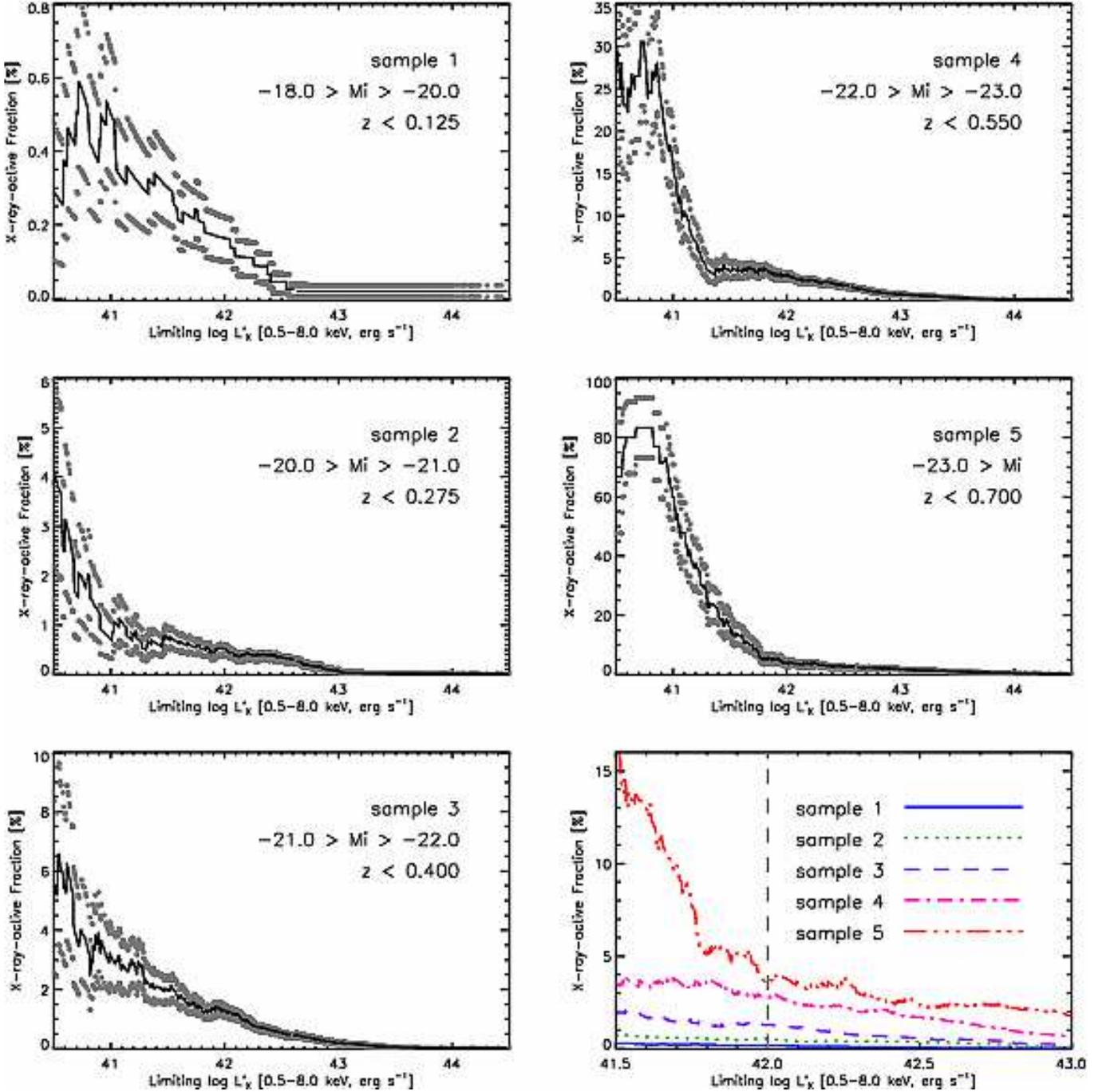}
\caption{The first five panels show the X-ray-active fraction as a
  function of limiting broad-band X-ray luminosity for each of the
  five volume-limited samples described in Table \ref{samples} and
  shown in Fig. \ref{Mi_z}. The solid lines connect the fraction as
  determined for the limiting $L_X^{\prime}$ at each X-ray detection
  in the sample. The $1~ \sigma$ errors (determined via the $\beta$
  distribution; see \S \ref{beta}) appear at each point where an
  active fraction was calculated. The final panel contains the best
  value for the fraction for each sample, plotted on the same scale
  for ease of comparison: sample 1 (solid blue line), sample 2 (dotted
  green line), sample 3 (dashed purple line), sample 4 (dot-dashed
  magenta line), and sample 5 (triple dot-dashed red line). The
  dashed gray line in this last panel denotes the limiting X-ray
  luminosity, log $L_X = 42$, that we associate with AGN; at this
  limit the AGN fractions are $F_{AGN,1} = 0.16\pm{0.06}$\%,
  $F_{AGN,2} = 0.50\pm{0.11}$\%, $F_{AGN,3} = 1.27\pm{0.18}$\%,
  $F_{AGN,4} = 2.85\pm{0.39}$\%, and $F_{AGN,5} = 3.80\pm{0.92}$\% for
  samples 1-5, respectively (see also Table \ref{agnfrac_tab}). \\
  (A color version of this figure is available in the online journal.)}
\label{agnfrac_fig}
\end{figure*}
%-------------------------------------------------------------

%--------------------------FIGURE 11---------------------------
\begin{figure}[ht]
\includegraphics[width=0.5 \textwidth]{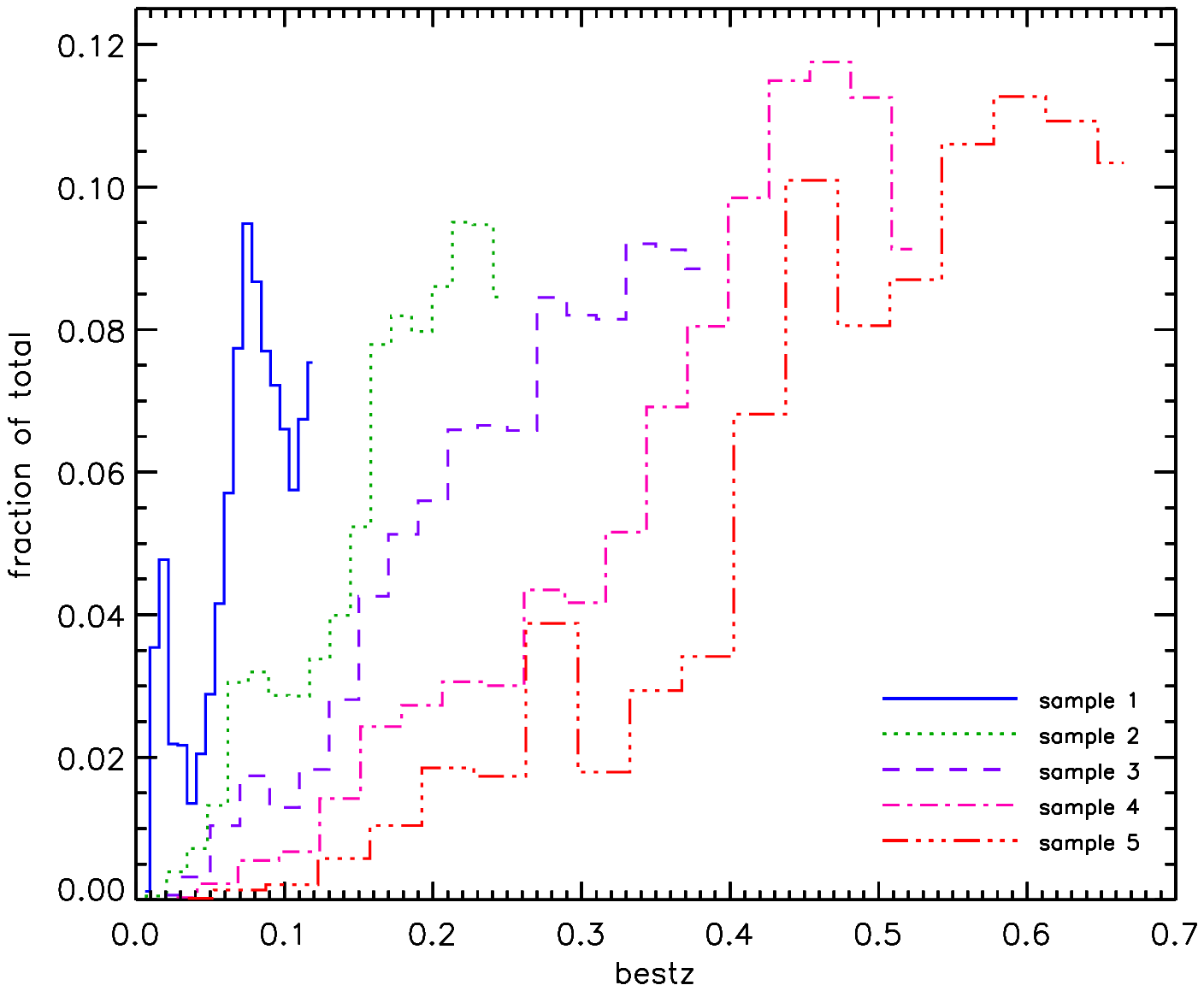}
\caption{Distribution of \bestz\ for each of the five samples
described in Table \ref{samples} and in Fig. \ref{Mi_z} (line styles
are as in the bottom, right panel of Fig. \ref{agnfrac_fig}). \\
(A color version of this figure is available in the online journal.)}
\label{multi_z_hist}
\end{figure}
%-------------------------------------------------------------

Figure \ref{agnfrac_fig} shows the X-ray-active fraction as a function
of $L_X^{\prime}$, where the luminosity threshold is set to the
measured $L_X$ for each of the X-ray-detected sources; panels 1--5
show the fraction for the five volume-limited optical samples. Errors
on the fraction due to small number statistics, estimated via the
$\beta$ distribution (gray filled circles; see \S \ref{beta} below),
are largest at the lowest luminosity limits, where both the number of
X-ray detections and galaxies is small (see also
Fig. \ref{agnfrac_num} in the Appendix). The sixth panel shows the
active fraction over a narrower range of $L_X^{\prime}$ from all five
samples on the same vertical scale (sample 1: blue solid line; sample
2: green dotted line, sample 3: purple dashed line; sample 4: magenta
dot-dashed line; sample 5: red triple dot-dashed line). The dashed
gray line denotes an X-ray luminosity limit of $L_X^{\prime} =
10^{42}$~erg\,s$^{-1}$, above which we consider our fraction a proper
AGN fraction (see \S \ref{agn_frac}), and below which the fraction may
be impacted by a mix of stellar emission (principally from X-ray
binaries) and hot ISM in the host galaxies.

The redshift distributions of the five optically-complete samples
differ substantially from one another, see Fig. \ref{multi_z_hist}.
Sample 1 (at the lowest redshifts) shows two peaks, one corresponding
to a large number of spectroscopic redshifts at $z=0.02$ and the other
at $z=0.075$; sample 2 peaks at $z=0.23$; sample 3 at $z=0.35$; sample
4 at $z=0.46$; and sample 5 at about $z=0.6$. As a result, the increase
in X-ray-active fraction between samples 1 and 5 may be due either to
a trend with $M_i$, with $z$, or with both. We discuss these
dependencies in \S \ref{agn_frac} and \S \ref{discussion}.

\subsection{Error Estimation via the $\beta$ Distribution}
\label{beta}

\begin{figure}[t]
\includegraphics[width=0.47 \textwidth]{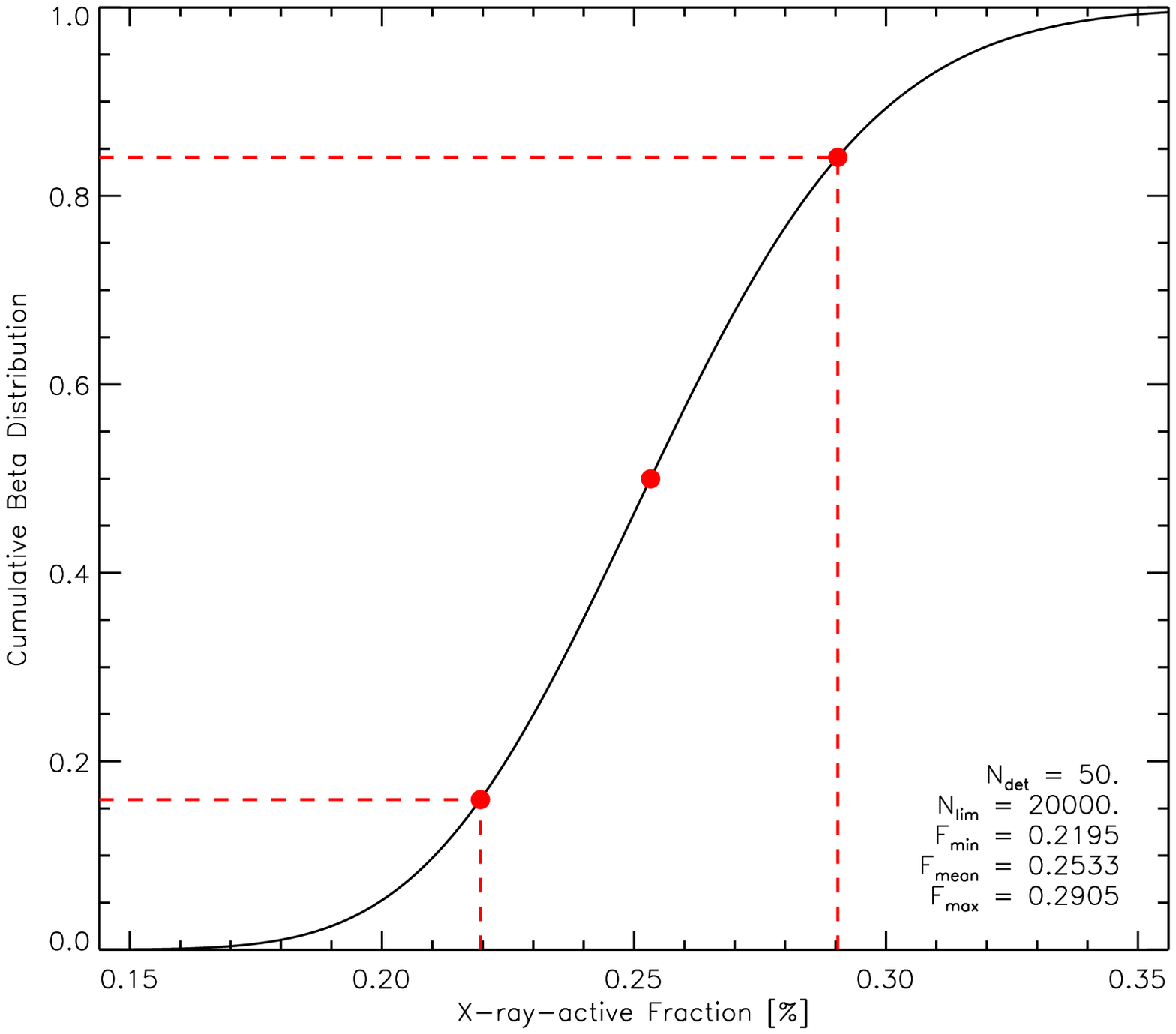}
\caption{The cumulative $\beta$ distribution function (black solid
  line) for a toy example where the number of detections, $N_{det}$,
  equals 50 and the number of X-ray observed galaxies, the denominator
  $N_{lim}$, is 20,000. For these inputs, the mean fraction is
  0.25\%. The mean value of the CDF, $F_{mean}$, is marked with a
  large, filled circle, as are the slightly asymmetric $1~\sigma$
  errors (also marked with dashed lines), corresponding to $F_{min}$
  and $F_{max}$.}
\label{beta_fig}
\end{figure}

We face small number statistics throughout this study, particularly
when we bin our modest number of X-ray-detections further by mass,
redshift, color, \etc\ Statistics used to estimate errors on numbers
of events, \eg, the Poisson distribution or the \cite{Gehrels86}
statistic, are not suitable for estimating errors on \emph{fractions}.
When it comes to estimating uncertainties on a fraction, one should
use the binomial distribution, or its conjugate prior, the $\beta$
distribution \citep{Evans00}. Given a fraction, $F$, and a number of
trials, $N$, the binomial distribution describes the probability of
different outcomes. Given the outcome of such a trial, the $\beta$
distribution describes the probability distribution for the fraction
itself. Hence, we determine both the X-ray-active (or AGN) fraction
and the uncertainties on the fraction using the regularized incomplete
$\beta$ distribution (see Equation \ref{beta_eqn} and
Fig. \ref{beta_fig}).

The $\beta$ distribution is continuous, yields a fraction that is well
behaved in the extremes where either the numerator or the denominator
(or both) are zero, and properly accounts for a numerator and
denominator that are not drawn from independent samples, \ie\ where
one is a subset of the other, as is the case here. We assign a
probability distribution to the fraction, $P_{\beta}$ (defined on the
interval [0, 1]), given the measured values $\alpha = N_{det} + 1$ and
$\beta = N_{lim} - N_{det} + 1$. For each input $N_{det}$ and
$N_{lim}$, we calculate a cumulative distribution function (CDF) via
the $\beta$ distribution and choose the mean of the CDF as the best
value for the X-ray-active fraction, such that
\begin{equation}
F_{mean} = {{\alpha} \over {\alpha + \beta}} = {{N_{det} + 1} \over
  {N_{lim} + 2}}~. 
\label{beta_eqn}
\end{equation}

\noindent{The asymmetric $1~\sigma$ errors\footnote{We take $1~\sigma
= 0.34134$.} are calculated based on the median value of $F$
(evaluated where $P_{\beta} = 0.5$), and defined such that $F_{max}$
corresponds to $P_{\beta}(0.5) + 1~\sigma$ and $F_{min}$ corresponds
to $P_{\beta}(0.5) - 1~\sigma$ (see Fig. \ref{beta_fig} for an
example). Throughout the remainder of this manuscript (tables
excepted), we specify the asymmetric $1~\sigma$ errors \emph{only} in
cases where the upper and lower errors differ by more than 10\%; in
all other cases we quote the larger of the two. (That is, if we have
$F\pm^X_Y$, we will quote $F\pm max(X,Y)$, unless $X$ and $Y$ differ
by more than 10\%.) The cumulative X-ray-active fractions and their
$\beta$ distribution errors, for a set of seven useful X-ray
luminosity thresholds, appear in Table \ref{agnfrac_tab}.}

%---------------------------FIGURE 12-------------------------
%-------------------------------------------------------------

%---------------------------TABLE 3---------------------------
%-------------------------------- AGN FRACTION TABLE ----------------------------------

\begin{deluxetable}{lccccccc} 
\tabletypesize{\scriptsize} 
\tablewidth{0pt}
\tablecaption{X-ray Active Fraction} 
\tablehead{
\colhead{} & 
\colhead{} & 
\colhead{} & 
\colhead{} & 
\colhead{} & 
\multicolumn{3}{c}{$F$ (\%)} \\
\cline{6-8}
\colhead{log $L_X^\prime$} &
\colhead{N$_{det}$} & 
\colhead{N$_{lim}$} & 
\colhead{N$_{det,drop}$} & 
\colhead{N$_{lim,drop}$} & 
\colhead{min} & 
\colhead{mean} & 
\colhead{max} }

\startdata
\multicolumn{8}{l} {Sample 1: -18 $\ge M_i >$ -20, bestz $\le$ 0.125} \\ % (29 detections, 5028 limits)} \\ 
\cline{1-8}
41.0      &    7   &   1587    &     7   &   3441  &   0.33   &   0.50   &   0.68 \\
41.5      &    8   &   2974    &     3   &   2054  &   0.20   &   0.30   &   0.40 \\
{\bf 42.0} & {\bf   6}   & {\bf  4336}    &  {\bf   0}   &  {\bf  692}  &  {\bf 0.10}   &  {\bf 0.16}   &  {\bf 0.22}\\
42.5      &    1   &   4930    &     0   &     98  &   0.01   &   0.04   &   0.07 \\
43.0      &    0   &   5026    &     0   &      2  &   0.00   &   0.02   &   0.04 \\
43.5      &    0   &   5028    &     0   &      0  &   0.00   &   0.02   &   0.04 \\
44.0      &    0   &   5028    &     0   &      0  &   0.00   &   0.02   &   0.04 \\
\cline{1-8}
\multicolumn{8}{l} {Sample 2: -20 $\ge M_i >$ -21, bestz $\le$ 0.275} \\ % (88 detections, 10949 limits)} \\
\cline{1-8}
41.0      &    2   &    420    &    77   &  10529  &   0.33   &   0.71   &   1.10 \\
41.5      &   11   &   1629    &    57   &   9320  &   0.53   &   0.74   &   0.94 \\
{\bf 42.0} & {\bf  21}   & {\bf  4415}    &  {\bf  30}   &  {\bf 6534}  &  {\bf 0.39}   &  {\bf 0.50}   &  {\bf 0.60}\\
42.5      &   24   &   7930    &     7   &   3019  &   0.25   &   0.32   &   0.38 \\
43.0      &    7   &  10240    &     0   &    709  &   0.05   &   0.08   &   0.11 \\
43.5      &    0   &  10903    &     0   &     46  &   0.00   &   0.01   &   0.02 \\
44.0      &    0   &  10949    &     0   &      0  &   0.00   &   0.01   &   0.02 \\
\cline{1-8}
\multicolumn{8}{l} {Sample 3: -21 $\ge M_i >$ -22, bestz $\le$ 0.400} \\ % (239 detections, 17984 limits)} \\
\cline{1-8}
41.0      &    8   &    280    &   208   &  17704  &   2.16   &   3.19   &   4.22 \\
41.5      &   21   &   1123    &   164   &  16861  &   1.55   &   1.96   &   2.36 \\
{\bf 42.0} & {\bf  48}   & {\bf  3860}    &  {\bf  90}   &  {\bf14124}  &  {\bf 1.09}   &  {\bf 1.27}   &  {\bf 1.45}\\
42.5      &   44   &   9047    &    42   &   8937  &   0.42   &   0.50   &   0.57 \\
43.0      &   23   &  14507    &     8   &   3477  &   0.13   &   0.17   &   0.20 \\
43.5      &    6   &  17343    &     1   &    641  &   0.03   &   0.04   &   0.06 \\
44.0      &    0   &  17949    &     0   &     35  &   0.00   &   0.01   &   0.01 \\
\cline{1-8}
\multicolumn{8}{l} {Sample 4: -22 $\ge M_i >$ -23, bestz $\le$ 0.550} \\ % (416 detections, 22095 limits)} \\
\cline{1-8}
41.0      &   13   &     83    &   378   &  22012  &  12.49   &  16.47   &  20.45 \\
41.5      &   13   &    395    &   356   &  21700  &   2.61   &   3.53   &   4.44 \\
{\bf 42.0} & {\bf  52}   & {\bf  1857}    &  {\bf 282}   &  {\bf20238}  &  {\bf 2.47}   &  {\bf 2.85}   &  {\bf 3.24}\\
42.5      &   98   &   6000    &   155   &  16095  &   1.49   &   1.65   &   1.81 \\
43.0      &   86   &  13209    &    44   &   8886  &   0.59   &   0.66   &   0.73 \\
43.5      &   44   &  19154    &     2   &   2941  &   0.20   &   0.23   &   0.27 \\
44.0      &    4   &  21732    &     0   &    363  &   0.01   &   0.02   &   0.03 \\
\cline{1-8}
\multicolumn{8}{l} {Sample 5: $M_i <$ -23, bestz $\le$ 0.700} \\ % (323 detections, 13628 limits)} \\
\cline{1-8}
41.0      &   11   &     18    &   311   &  13610  &  49.10   &  60.00   &  70.87 \\
41.5      &   13   &     84    &   291   &  13544  &  12.34   &  16.28   &  20.23 \\
{\bf 42.0} & {\bf  15}   & {\bf   419}    &  {\bf 260}   &  {\bf13209}  &  {\bf 2.88}   &  {\bf 3.80}   &  {\bf 4.72}\\
42.5      &   42   &   1842    &   199   &  11786  &   1.98   &   2.33   &   2.68 \\
43.0      &   96   &   5499    &    88   &   8129  &   1.59   &   1.76   &   1.94 \\
43.5      &   94   &  10085    &    15   &   3543  &   0.85   &   0.94   &   1.04 \\
44.0      &   51   &  12875    &     2   &    753  &   0.35   &   0.40   &   0.46 \\
\enddata

%\vspace{-7mm}
\tablecomments{The first column gives the broad-band (0.5-8.0 keV)
X-ray luminosity threshold. The second and third columns contain the
number of detections and limits that were used to calculate the mean
X-ray-active fraction, $F_{mean}$; columns four and five contain the
number of objects that did not pass the X-ray luminosity cut (details
in the Appendix). $F_{min}$ and $F_{max}$ are the $1-\sigma$ upper and
lower limits on the X-ray-active fraction (see \S \ref{beta}). The
bold rows indicate the AGN fraction, \ie\ the fraction for log
$L_X^{\prime} = 42$.}
\label{agnfrac_tab}
\end{deluxetable}
%---------------------------------------------------------------------------

%\label{agnfrac_tab}
%-------------------------------------------------------------

\subsection{The AGN Fraction}
\label{agn_frac}

In this section and the discussion that follows, we differentiate
between the X-ray-active fraction, which can be defined for an
arbitrary $L_X^{\prime}$ threshold, and the AGN fraction,
$F_{AGN}$. For our purposes, the AGN fraction will refer only to the
cases where the limiting X-ray threshold is $L_X^{\prime} =
10^{42}$~erg\,sec$^{-1}$, a typical cutoff for AGN activity
\citep{Zezas98}. For each of our volume-limited samples, we measure
the following AGN fractions (see also Table \ref{agnfrac_tab}, bold
entries): $F_{AGN,1} = 0.16\pm{0.06}$\%, $F_{AGN,2} =
0.50\pm{0.11}$\%, $F_{AGN,3} = 1.27\pm{0.18}$\%, $F_{AGN,4} =
2.85\pm{0.39}$\%, and $F_{AGN,5} = 3.80\pm{0.92}$\%.

%--------------------------FIGURE 13---------------------------
\begin{figure*}[htb]
\includegraphics[width=1.0 \textwidth]{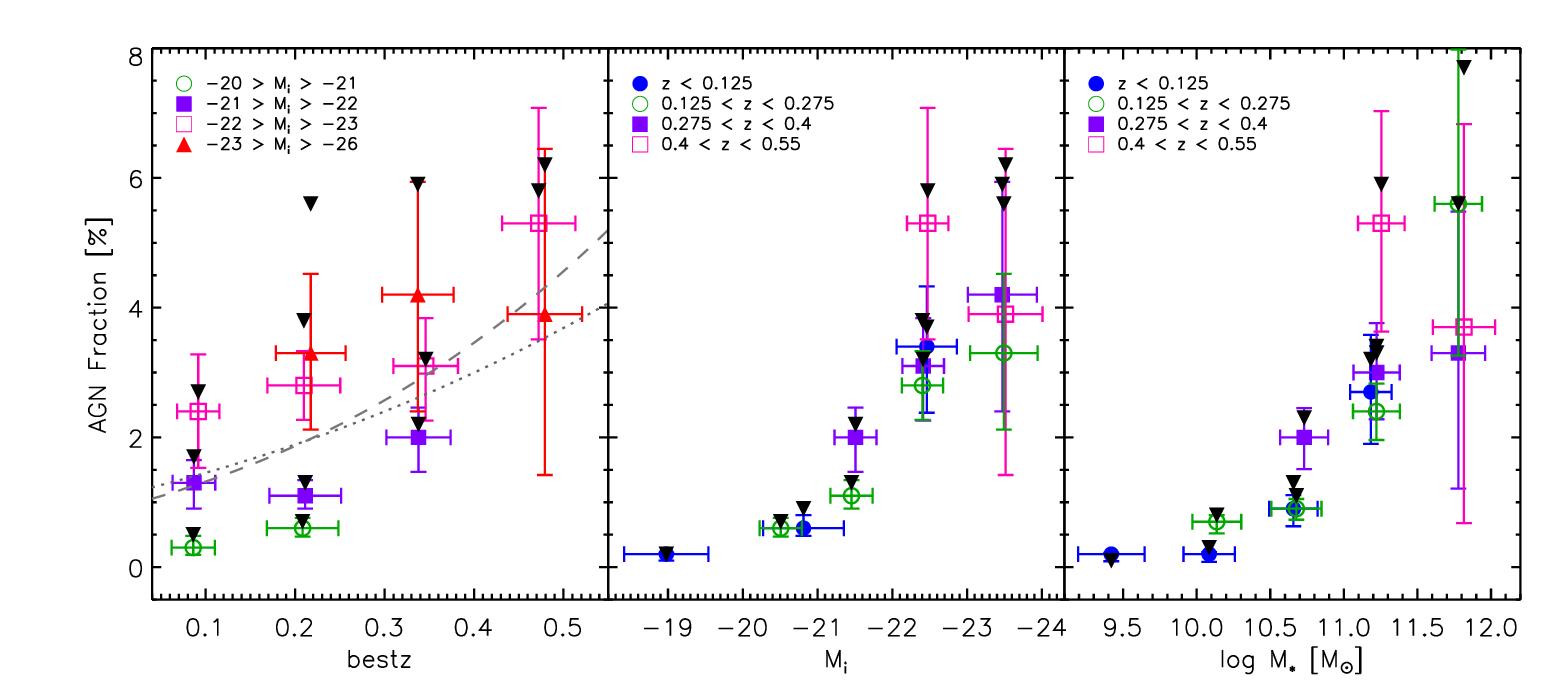}
\caption{AGN fraction, \ie\ $L_X^{\prime} = 10^{42}$~erg\,sec$^{-1}$,
in bins of redshift ({\it left}), absolute $i$-band magnitude ({\it
center}), and mass ({\it right}). Upper limits on the fractions are
plotted as solid, downward-facing black triangles (see \S \ref{bias}
for details). Two common fits for the evolution of the X-ray
luminosity function of X-ray-detected AGN, normalized to the mean AGN
fraction in the four $M_i$ bins at $z \sim 0.21$, are shown in the
left-hand panel: pure luminosity evolution $\propto (1+z)^3$ (gray
dotted line), and luminosity-dependent density evolution $\propto
(1+z)^4$ (dashed gray line). The Spearman rank correlation test
indicates that the active fraction correlates most strongly with
galaxy mass (93\% confidence; 96\% if the $0.4 < z < 0.55$ redshift
bin is ignored). There is also a correlation with $M_i$ (91\% and 96\%
confidence, respectively). A weaker correlation with \bestz\ is
detected at only the 76\% confidence level. (A color version of this
figure is available in the online journal.)}
\label{trends}
\end{figure*}
%-------------------------------------------------------------

To test for evolution in the AGN fraction with \bestz, and/or a change
with $M_i$ or galaxy mass, we further subdivide our five
volume-limited samples, re-calculate the AGN fraction in these finer
bins, and perform a Spearman rank correlation analysis (including
errors) to assess the significance of any trends. We first sort our
data into four absolute magnitude bins, $-20 \ge M_i>-21$, $-21 \ge
M_i>-22$, $-22 \ge M_i>-23$, $-23 \ge M_i>-24$, and simultaneously fit
all twelve of the data points shown in the left-most panel of
Fig. \ref{trends}. We find a marginally significant trend at the
$\sim$76\% confidence level between the AGN fraction and \bestz. In
four redshift bins, $z<0.125$, $0.125<z<0.275$, $0.275<z<0.4$, and
$0.4<z<0.55$, we find that the AGN fraction correlates strongly with
$M_i$, at the 91\% confidence level (96\% if the $0.4<z<0.55$ redshift
bin is ignored; Fig. \ref{trends}, middle panel). For the same
redshift bins, the correlation with galaxy mass is similar, but
slightly stronger, $\sim$93\% and 96\% with and without the
highest-redshift bin (Fig. \ref{trends}, right panel).

These trends reveal a field AGN fraction that increases toward
brighter absolute magnitudes and higher masses, as might be expected
from cosmological models wherein the merger of massive galaxies fuels
both accretion onto a central SMBH and star formation. The weaker
evolution of the AGN fraction with redshift indicates that the
mechanism for black hole fueling may be changing, \eg, shifting from
high luminosity merger-driven activity to fueling via tidal
interactions, bars, dynamical friction, or other lower luminosity
drivers \citep[see][for a review]{Martini04}. In such a scenario, the
lower AGN fraction may reflect the evolution of the AGN X-ray
luminosity function, which is characterized by a decrease in accretion
luminosity toward the present day. In the left-most panel of
Fig. \ref{trends} we show two common fits to the AGN X-ray luminosity
function \citep{Ueda03,Barger05a,Hasinger05}, normalized to the mean
AGN fraction in the four $M_i$ bins at $z \sim 0.21$: pure luminosity
evolution $\propto (1+z)^3$ (gray dotted line), and
luminosity-dependent density evolution $\propto (1+z)^4$ (dashed gray
line). We explore the possibility of an evolving AGN fraction further
in \S \ref{discussion}.

\subsection{Redshift Monte Carlo Simulations}
\label{mc}

As discussed in \S \ref{beta}, small number statistics for
X-ray-detected sources in our volume-limited bins place a basic limit
on our ability to accurately determine the X-ray-active
fraction. However, our use of photometric redshifts for the majority
of the galaxies ($\sim 97$\%) in our X-ray-observed sample introduces
another potentially large source of error. Photometric redshift errors
are of order 3\%, but can be as high as 10\% at fainter apparent
magnitudes \citep{Csabai03,Oyaizu08}. Meanwhile, photometric redshifts
for quasars can experience catastrophic failures due to aliasing
\citep{Budavari01,Richards01}. Hence, photometric redshift errors
dominate the uncertainty on our volume-limited samples. To estimate
the impact of the photo-$z$ uncertainties on our active fraction, we
perform two Monte Carlo (MC) simulations: a ``full'' MC and a
``quick'' MC.

\vspace{1mm}
{\noindent {\bf Full MC:} In the full Monte Carlo simulation, the $1~\sigma$
error for each object not associated with a QSO photometric redshift
is multiplied by a normally-distributed random number selected from a
distribution with a mean of zero and a standard deviation of one. This
quantity is added to the best redshift, yielding a new \bestz.}

Since single-valued quasar photometric redshifts, derived using broad
band filters, suffer from aliasing in certain redshift regimes they
are better described by their full redshift probability distribution
function \citep[see, \eg,][and references
therein]{Myers09,Richards09}. Thus, for objects where we have assigned
a QSO photometric redshift we use the full redshift PDF, obtained
using the nearest neighbor approach of \cite{Ball08}, to calculate the
new \bestz. We also run the MC tests using the QSO $1~\sigma$ errors
described in \S \ref{redshift}, instead of the full PDF. The results
of those tests show excellent agreement with the MCs that use the QSO
PDF. Thus, we do not attempt to acquire full redshift PDFs for all of
the objects in our sample.

We use the new \bestz\ values for non-QSO and QSO objects to determine
new luminosity distances, k-corrections, and absolute magnitudes via
{\tt kcorrect}. We then pull five volume-limited samples analogous to
those used for our primary analysis using the new \bestz\ and its
associated $M_i$. We adopt the same seven X-ray luminosity thresholds
and calculate new X-ray-active fractions for each.

The above constitutes a single iteration of the full MC. The process
is computationally expensive; determining the spectral template match
and k-correction for each of $\sim$ 100,000 objects is the largest
computational task. Thus, we perform only 100 iterations of the full
MC, which result in 100 possible X-ray-active fractions for each
volume-limited sample, at each of seven X-ray luminosity limits. We
take the median and standard deviation of these fractions and use them
to estimate the uncertainty contributed by the photo-$z$ errors. We
also use the full MC to estimate the representative error bars shown
in Figs. 6--9.

\vspace{1mm} {\noindent {\bf Quick MC:} In our quick Monte Carlo
simulation, new \bestz\ values are derived as in the full MC and new
luminosity distances are then calculated. We assume that a small
change in the redshift will result in only a small change in the
k-correction and so leave the k-correction unaltered. Hence, the new
absolute magnitude is simply the original $M_i$ weighted by the ratio
of the luminosity distances: $M_{i,new} = M_i - 5~{\rm
log}~(D_{L,new}/D_L)$.  We perform this quick MC initially for 100
iterations and make a direct comparison to our full MC. We then
perform 1000 iterations to achieve more robust statistics. Here again,
we derive 100 (or 1000) X-ray-active fractions and compute a median
fraction and standard deviation for each volume-limited sample at each
of seven X-ray luminosity thresholds.}

The errors that result from the redshift MCs are, in general, smaller
than the $\beta$ distribution errors, though in some cases they are of
comparable magnitude. For example, if we adopt an X-ray threshold of
$L_X^{\prime} = 10^{42}$~erg\,s$^{-1}$ for sample 3, the 100-iteration
full MC gives a median fraction of $1.25\pm{0.12}$\%, the
100-iteration quick MC yields a median of $1.25\pm{0.11}$\%, and the
1000-iteration quick MC gives a median fraction of
$1.24\pm{0.11}$\%. These compare to a $\beta$ distribution result of
$1.27\pm{0.18}$\%, where the $\beta$ distribution errors arise from
small numbers of X-ray detected galaxies, instead of photo-$z$
uncertainties.  The $\beta$ distribution and MC errors could be
combined in quadrature; or as a conservative estimate, the errors in
Table \ref{agnfrac_tab} might simply be increased by a factor of
$\sqrt{2}$ to take photometric redshift errors into account.

\subsection{Selection Effects and Sample Bias}
\label{bias}

As discussed in \S \ref{redshift}, we have included a large number,
($\sim$27 thousand) of red, $(u-g) > 0.6$, point sources in our galaxy
sample to insure that we do not bias our AGN fraction at faint
apparent magnitudes (and at high-redshift) where morphological
criteria are less reliable, and to include all possible QSOs. As a
result, the denominator of our fractions is likely to be contaminated
by stars and the AGN fractions reported here must be taken as a lower
limit to the ``true'' AGN fraction.

To quantify the magnitude of the stellar contamination, we return to
SDSS star-galaxy separation \citep{Stoughton02}. We assume that all of
the point sources with red colors ($u-g>0.6$) \emph{are} actually
stars and remove these 27,958 objects along with their X-ray
counterparts (120 detections) from our galaxy sample. We re-run our
analysis and find mean AGN fractions at $L_X^{\prime} =
10^{42}$~erg\,sec$^{-1}$ of: $0.16\pm{0.08}$\% (sample 1),
$0.61\pm{0.14}$\% (sample 2), $1.53\pm{0.33}$\% (sample 3),
$3.45\pm{0.46}$\% (sample 4), and $5.71\pm{1.37}$\% (sample 5).  Since
some of these excluded objects are certainly galaxies, and the number
hosting X-ray-emitting AGN is likely to be smaller than the passive
population, we consider these fractions to be upper limits, \ie\ we
will have eliminated more legitimate galactic objects from the
denominator than the numerator. We plot these ``maximum'' AGN
fractions in Fig. \ref{trends} (filled, black, downward-facing
triangles) and note that, with few exceptions, they agree with our
original AGN fractions within the errors.

We also consider the possibility that the combination of ChaMP and
SDSS spectroscopy (which includes known X-ray sources in its
targeting algorithms) introduces a bias in our redshifts, leading to
greater spectroscopic completeness in the X-ray-detected sample than
in the X-ray-observed sample. This is, at worst, a second-order effect
impacting only the precision of the redshifts since we include
galaxies in our sample whether or not they have spectroscopy. One
exception is along the stellar locus. In regions of the color-color
diagram where we have eliminated objects based on their photometric
redshifts and colors (see \S 5.5), we do \emph{not} eliminate
spectroscopically-confirmed galaxies. Thus, we may have an
over-representation of objects in these particular color regimes due
to existing ChaMP and SDSS spectroscopy.

Pollution of the host galaxy light by a central AGN or active star
formation is another source of bias in our galaxy sample. Though we
are interested in locating active black holes in nearby galaxies,
these same sources may contaminate the spectral energy distribution
(SED) of the host galaxy and thus impact the photometric redshift
algorithms and {\tt kcorrect} spectral template matching. There are
several possible approaches to this problem: (1) in principle it is
possible to correct the host galaxy SED by extrapolating an AGN
powerlaw spectrum, \eg, normalized to the measured X-ray luminosity,
into the optical and then subtracting this component. This method is
not likely to be efficient for our sample because only a small number
($\sim 1.5$\%) of our galaxies have X-ray detections. Also, a
non-negligible fraction of X-ray AGN have the optical SEDs of normal
galaxies \citep{Georg05,Horn05,Kim06,Civano07,Cocchia07}, so
subtraction of a canonical normalized SED could introduce other
errors.  (2) Where spectra are available, the AGN contribution can be
estimated and removed from the host galaxy SED. With limited
spectroscopic coverage ($\sim 2.5$\% of our galaxies have spectra),
this too is an imperfect approach for these data. (3) Lastly, fitting
of high-quality imaging data can successfully deconvolve the AGN point
source from the host galaxy light. However, since the large majority
of our objects are faint such an approach is not feasible.

%--------------------------FIGURE 14---------------------------
\begin{figure}[htb]
\includegraphics[width=0.5 \textwidth]{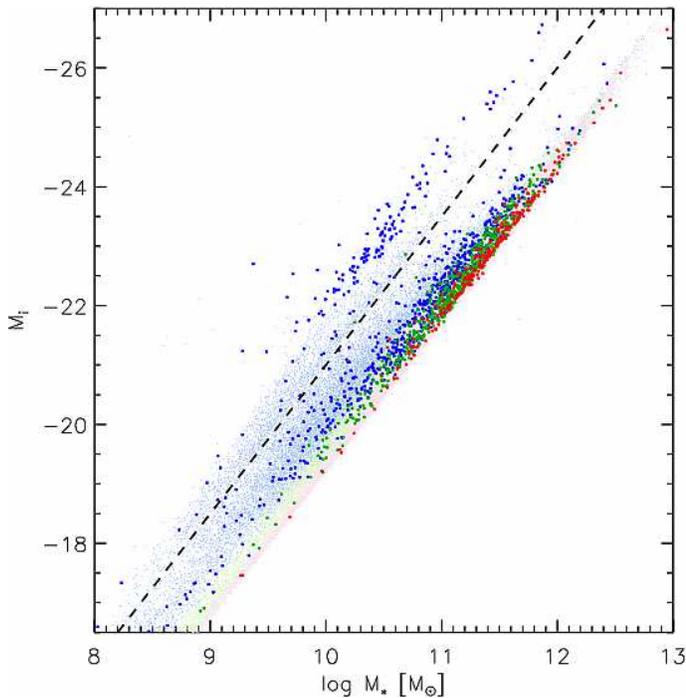}
\caption{Absolute $i$-band magnitude vs. log $M_{\star}$ for the
  limits (small points) and detections (stars) in the Full galaxy
  sample. The symbols are colored according to the $(u-r)_0$ criteria
  described in \S \ref{field_frac}. The strong correlation between
  magnitude and mass is expected as more massive galaxies tend to be
  more luminous. In addition, both are calculated here from the fit
  performed by the {\tt kcorrect} tool \citep{Blanton07}. There is a
  population of objects that define a ``second'' locus or whose $M_i$
  or mass determinations are flawed. We differentiate these objects by
  eye (dashed line) and consider the possibility that these contain a
  central AGN that dominates the host galaxy SED (see \S \ref{bias}).}
\label{mass_mi}
\end{figure}
%-------------------------------------------------------------

We instead make a qualitative assessment of how an AGN contribution to
our host galaxy SED would bias our fractions. In Fig. \ref{mass_mi},
we plot the $i$-band absolute magnitude versus the stellar mass of our
galaxy sample (small points) and the X-ray-detected subset (stars),
color-coded according to the following $(u-r)_0$ color sequences:
blue: $(u-r)_0 \le 1.8$, green: $1.8 < (u-r)_0 < 2.6$, and red:
$(u-r)_0 \ge 2.6$. Since more massive galaxies tend to be more
luminous, it is not remarkable that these quantities show a strong
correlation. In our analysis, the galaxy magnitudes and masses are
derived using the same fit to the models implemented in the
\cite{Blanton07} {\tt kcorrect} tool, further insuring a correlation
between the two. We note a distinct second population of blue objects
with apparently higher masses and brighter luminosities; we demarcate
this ``second sequence'' by eye (as indicated by the black dashed line
in Fig. \ref{mass_mi}). Using this simple cut, we identify 3,023
X-ray-observed and 156 X-ray-detected galaxies where AGN contamination
may be an issue. These objects are indicated in Figs. \ref{mi_lx_z}
and \ref{mass_mi} (black points [observed] and black open circles
[detected]). Where we have only upper limits to the X-ray luminosity,
the galaxies' very blue colors are our primary evidence that they host
an AGN. To be conservative, here we assume that all of these galaxies
do host an AGN whose X-ray emission is either weak or obscured, but
whose optical continuum substantially contaminates the colors.

We might explain this second sequence by speculating that the
k-correction and derived $M_i$ and stellar mass for these objects is
incorrect. For example, we might estimate, based on Fig. \ref{kcorr},
that the derived $M_i$ for these objects is too faint by about one
magnitude. Hence, for an object at $M_i \sim -23$, we would correct up
by one magnitude to $M_i \sim -24$. An object on the primary sequence
at this magnitude corresponds to a stellar mass of log $M_{star} \sim
11.8$, very reasonable for an AGN host.

Mischaracterization of the k-correction is also likely to bias our AGN
fraction as a function of color, yielding an artificially high
fraction in the blue sequence. We have, however, tested the impact of
these ``second sequence'' objects on our global AGN fraction as a
function of $M_i$ and $bestz$ and found that the fraction is robust to
their inclusion (or exclusion).

Among these potentially contaminated objects, we associate 121 ($4$\%)
with QSO photometric redshifts --- this compares to $\sim 0.3$\% of
objects with QSO photometric redshifts in the Full sample. Another 83
have spectra, of which 66 have secure spectroscopic types: there are
50 broad line AGN (BLAGN) identified via ChaMP spectroscopic
follow-up, a single ChaMP narrow emission line galaxy, and 15 SDSS
MPA/JHU narrow emission line galaxies. Thus, a significant fraction of
the objects along this second locus are indeed associated with active,
starforming galaxies and/or likely AGN. However, since these objects
lie preferentially outside of the limits of our five volume-limited
samples --- only one X-ray-detection and fewer than 300 galaxies
across all five samples appear in our AGN fraction calculations --- we
do not recalculate $F_{AGN}$ without these objects. We do, however,
caution that AGN activity may bias our derived host galaxy properties
in a small number of cases and in the highest redshift, highest
absolute magnitude sample, in particular.

White dwarfs are another possible contaminant in our galaxy sample.
Their blue colors and X-ray brightness can cause them to masquerade as
AGN, particularly if they have been incorrectly assigned a
cosmological photometric redshift. However, since we removed objects
with significant proper motions, we anticipate that these will not
contribute more than an object or two to our numerator and thus will
not seriously bias our fractions.

\section{Discussion}
\label{discussion}

In this section we compare our results with previous AGN fraction
measurements in the field and in groups/clusters. We also expand our
analysis of the AGN fraction as a function of host galaxy
properties. To extract Vega UBVRI absolute magnitudes, k-corrections,
and masses, we re-run the {\tt kcorrect} tool on our Full galaxy
sample \citep{Bessell90,Blanton07}; these photometric bands are useful
for comparisons with other studies in the literature.

\subsection{Comparison to Previous Field Fractions}
\label{field_frac}

\cite{Page01} combines the present-day optical luminosity function of
galaxies with the X-ray luminosity function of Seyfert 1s to calculate
an expected AGN fraction. To reproduce the local type 1 AGN X-ray
luminosity function for a plausible mix of AGN hosts, including both
early- and late-type galaxies, he finds a Seyfert 1 fraction of about
1\% at $z \sim 0$. This compares favorably to our direct measurement
using intermediate-luminosity galaxy sample 3 ($-21 > M_i > -22$) at
low-redshift, for which we find an AGN fraction of $1.27\pm{0.18}$\%.

\begin{figure}[tb]
\includegraphics[width=0.5 \textwidth]{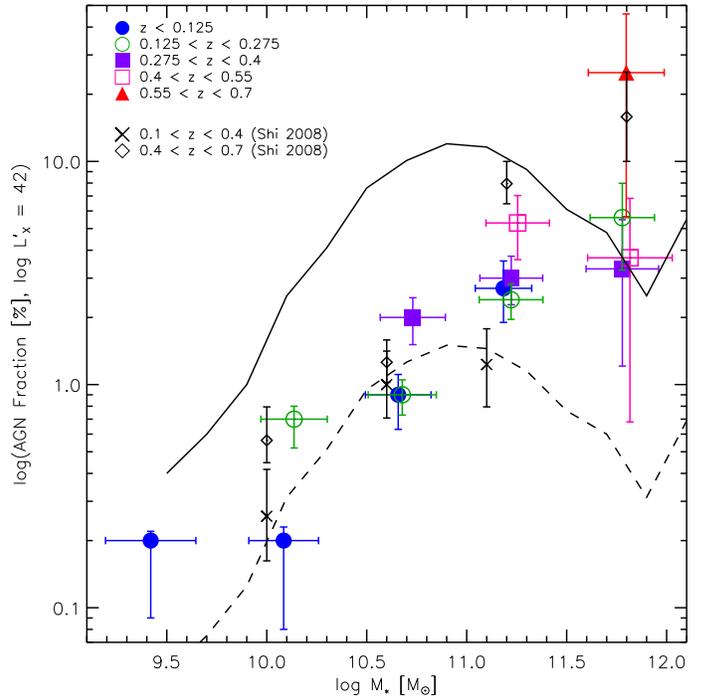}
\caption{Log of the AGN fraction (log $L_X^{\prime} = 42$, solid black
  line in the right panel of Fig. \ref{mi_lx_z}) versus stellar
  mass. Colored symbols indicate our AGN fractions for five redshift
  intervals ($z<0.125$, blue filled circles; $0.125 < z < 0.275$,
  green open circles; $0.275 < z < 0.4$, purple filled squares; $0.4 <
  z < 0.55$, magenta open squares; $0.55 < z < 0.7$, red filled
  triangles) from the mass-selected samples of Fig. \ref{mass_z} (left
  panel). The high- and low-redshift fractions from \cite{Shi08} are
  plotted as black symbols ($0.1 < z < 0.4$ as crosses, $0.4 < z <
  0.7$ as open diamonds). The solid line shows the trend of the SDSS
  powerful AGN fraction with the host stellar mass from
  \cite{Kauffmann03a}; the dashed line shows the same trend with the
  normalization decreased by a factor of 8
  \citep[following][]{Shi08}. \\
  (A color version of this figure is available in the online journal.)}
\label{mass_trend}
\end{figure}

From a compilation of deep X-ray surveys (including the \Chandra\ Deep
Fields, the \Chandra\ Large-Area Synoptic X-Ray Survey of Lockman Hole
[CLASXS], and the All-wavelength Extended Groth strip International
Survey [AEGIS]; covering $\sim 2~{\rm deg}^2$), \cite{Shi08} derive an
X-ray AGN fraction as a function of host stellar mass. We present the
AGN fraction versus host mass in Fig. \ref{mass_trend} for our
ChaMP/SDSS sample (colored symbols indicate our five redshift
intervals) and from \cite{Shi08} (black symbols; crosses for $0.1 < z
< 0.4$ and diamonds for $0.4 < z < 0.7$).  The fractions from these
studies agree to within $1~\sigma$ in all but one case, the highest
mass bin of their low-redshift sample, where the agreement is good to
within $1.5~\sigma$. We also include a comparison to the powerful AGN
fraction from an optical, spectroscopic study of SDSS galaxies
\citep[solid black line; ][]{Kauffmann03a}. \cite{Shi08} re-normalize
this relation by a factor of 8 to fit their data and account for
differences between the X-ray and optical selection functions, \eg,
emission-line-selected AGN samples are more complete (less sensitive
to obscuration) for low-mass galaxies at low-redshift than
X-ray-selected AGN \citep{Heckman05}. The \cite{Kauffmann03a} SDSS
powerful AGN fraction is measured for $z < 0.3$ and shows an increase
with stellar mass to $10^{11} M_{\odot}$ and then a sharp decrease
toward galaxies of higher masses. \cite{Shi08} interpret their
low-redshift sample as demonstrating a turn over (or flattening) in
the AGN fraction near $10^{11} M_{\odot}$, though no such inflection
is detected for their high-redshift sample. Neither our low- or
high-redshift samples show a turn over in the fraction, but instead
increase from the lowest-mass to the most massive galaxies in our
study.

%--------------------------FIGURE 15---------------------------
%-------------------------------------------------------------

We also compare our field AGN fraction to the zCOSMOS survey results
\citep{Silverman09a}, which utilize XMM-Newton observations of the
COSMOS fields. For a mass limit of log $M_{\star} > 10.4$ (black
dashed line in the right panel of Fig. \ref{mass_z}) we match the
redshift and X-ray luminosity intervals used by \cite{Silverman09a} in
their Fig. 11: $0.1 < z < 0.58$ and $0.58 < z < 1.05$, with $L_X >
10^{42.5}$~erg\,s$^{-1}$ (dashed black line in the right panel of
Fig. \ref{mass_z}). In the low redshift bin we have 186 X-ray
detections and 19,057 X-ray-observed galaxies; these yield an AGN
fraction of $0.98\pm{0.07}$\%. The zCOSMOS field fraction in this bin
is $1.9\pm{0.6}$\%. Hence, our field AGN fraction falls below the
zCOSMOS field fraction by $1.5~\sigma$. The difference between the
ChaMP/SDSS and zCOSMOS low-redshift results may reflect evolution in
the AGN fraction if the zCOSMOS sample is weighted toward
higher-redshift objects (in the low-redshift bin the ChaMP/SDSS sample
has an average redshift of $\langle z \rangle = 0.28$). In the high
redshift bin, we find 14 detections and 296 limits, which yield a
fraction of $5.03\pm{1.25}$\%, vs. the zCOSMOS value of
$3.7\pm^{0.6}_{0.5}$\%. However, we do not consider this second a
robust comparison because our sample is not complete beyond redshifts
of 0.4 for this mass cut (see Fig. \ref{mass_z}, left panel).

Evidence for evolution in the AGN fraction has been seen in other
recent work. \cite{Lehmer07} use the \Chandra\ Deep Fields to study
the AGN fraction in early-type galaxies (to $z \sim 0.7$) and find
evolution consistent with the $(1+z)^3$ pure luminosity evolution
(PLE) model frequently fit to the luminosity function of
X-ray-selected AGNs \citep[hereafter, the AGN
XLF; ][]{Ueda03,Barger05a,Hasinger05}. Similar behavior is also noted in
an earlier analysis of the stacked X-ray properties of early-type
galaxies in a $1.4~deg^2$ field in the NOAO Deep Wide-Field Survey
\citep{Brand05}. (Note, however, in a study of late-type galaxies,
\cite{Lehmer08} find no evidence for evolution of the AGN fraction
over the redshift range $0.1 < z < 0.8$.) In the left panel of
Fig. \ref{trends} we show curves for PLE ($\propto (1+z)^3$; gray
dotted line) and for luminosity-dependent density evolution (LDDE),
another popular model for the evolution of the AGN XLF ($\propto
(1+z)^4$; gray dashed line). These curves are intended to be
illustrative only and are thus arbitrarily normalized to the mean AGN
fraction in the four $M_i$ bins with $0.125 \le z < 0.275$. For the
$-22.0 \ge M_i > -23.0$ magnitude bin, we compare the AGN fraction in
the lowest and highest redshift intervals and find, $F_{AGN}(z=0.47) =
2.2\pm^{+2.5}_{-1.1} \times F_{AGN}(z=0.09)$; \ie\ though poorly
constrained, the change in this AGN fraction is consistent with either
$(1+z)^3$ or $(1+z)^4$ evolution.

%---------------------------TABLE 4---------------------------
%-------------------------------- MARTINI COMPARISON TABLE ----------------------------------

\begin{deluxetable*}{lcccccccc} 
\tabletypesize{\scriptsize} 
\tablewidth{0pt}
\tablecaption{Comparison to Cluster Fractions} 
\tablehead{\colhead{} & \multicolumn{2}{c}{log $L_X$} & \colhead{} & \colhead{} & \multicolumn{3}{c}{$F_{field}^a$ (\%)} & \colhead{} \\
\cline{2-3}\cline{6-8}
\colhead{$M_{R} <$} & \colhead{min} & \colhead{max} & \colhead{N$_{det}$} & \colhead{N$_{lim}$} & \colhead{min} & \colhead{mean} & \colhead{max} & \colhead{$F_{clust}^b$ (\%)} }

\startdata
-20   & 40.9 & 43.6   &  26 &  448 &  4.89 &  6.00 &  7.11 & 6.0 \\
-20   & 41.0 & \ldots &  28 &  642 &  3.69 &  4.50 &  5.31 & 4.9 \\
-20   & 42.0 & \ldots & 113 & 9588 &  1.08 &  1.19 &  1.30 & 1.0 \\
-21.3 & 41.0 & \ldots &  25 &  208 & 10.12 & 12.38 & 14.64 & 9.8 \\
-21.3 & 41.0 & \ldots &   4 &   46 &  6.14 & 10.42 & 14.71 & 5.5 \\
\enddata

\tablecomments{$^a$ The field fractions calculated from the ChaMP and SDSS survey data and described in the present work for $0.05 < z < 0.31$. $^b$ The first four cluster fractions come from \cite{Martini07} and also span $0.05 < z < 0.31$; the final comparison is to \cite{Sun07} clusters with $0.01 < z < 0.05$. Absolute R magnitudes are calculated via the SDSS transformations described in \S \ref{derived}. The \cite{Sun07} result in the table is a simple fraction determined from their values of 9 AGN and 163 galaxies. The mean AGN fraction for these inputs, calculated via the beta distribution, is 6.1\%.}
\label{martini_comp}
\end{deluxetable*}

%---------------------------------------------------------------------------

%\label{martini_comp}
%-------------------------------------------------------------

In the current work, we predicate our comparison to the AGN XLF on the
assumption that the underlying galaxy population does not evolve
substantially between redshifts of $z=0.5$ and the present day ---
indeed, we assume no evolution in our galaxy k-corrections. Thus, the
evolution of the AGN fraction is entirely attributed to the evolution
of the X-ray-detected AGN population. The assumption of zero evolution
for the galaxy population is almost certainly too simple, particularly
for late-type galaxies, which have been shown to fade by a magnitude
or more between $z \sim 1$ and the present
\citep{Lilly95,Wolf03,Faber07}. In fact, the lack of evolution of the
AGN fraction in late-types reported by \cite{Lehmer08}, may indicate
that the AGN and late-type galaxy populations co-evolve for $z < 1$.

\subsection{Comparison to Cluster and Group X-ray Fractions}
\label{clusters}

The impact of environment on the formation and evolution of accreting
supermassive black holes remains an important open question, \ie\ are
AGN preferentially located in field galaxies, in galaxy groups, or in
rich clusters? Early studies revealed a paucity of optically-luminous
AGN in clusters \citep{Dressler85}, compared to the field --- 7\%
vs. 31\% of galaxies had emission-line nuclei in clusters vs. the
field, and 1\% vs. 5\% harbored AGN. \cite{Giovanelli85} proposed
that galaxies in clusters lose cold gas from their disks by
interacting with the cluster core, leaving behind a cluster galaxy
population deficient in the cold gas necessary to fuel AGN. A smaller
frequency of AGN fueling events was also proposed --- in this
scenario, the merger of two gas-rich galaxies was invoked as the
mechanism for driving cold inflows toward a massive central black hole
\citep{Barnes92}. If galaxies in the cluster environment experience
fewer mergers due to their high velocity dispersions, less AGN
activity would result. These two scenarios might also work in tandem
to produce smaller AGN fractions in clusters. (Note that groups, which
have lower velocity dispersions than clusters, should have merger
rates closer to those in the field.)

The last decade has shown great advances in our understanding of the 
X-ray-selected population of AGN in clusters of galaxies
\citep{Martini02,Martini06,Martini07,Martini09}. Enabled by the
exceptionally high sensitivity and spatial resolution of \Chandra,
\cite{Martini02} and \cite{Martini07} showed that the fraction of
X-ray-selected AGN observable in clusters at $0.05 < z < 0.31$ was, in
fact, higher than previously reported and similar to the field, though
the field fraction has historically been poorly constrained. 

To compare our robust field X-ray AGN fractions to those reported for
clusters, we employ absolute R magnitudes, $M_R$, adopt a variety of
$L_X^{\prime}$ and $M_R$ limits, and measure the active fraction for
$0.05 < z < 0.31$ \citep[to match the cluster redshift range
of][]{Martini07}. These comparisons are summarized in Table
\ref{martini_comp}. Our field fractions for $M_R < -20$ are: (1)
$6.00\pm{1.11}$\% ($8 \times 10^{40} < L_X <
4\times10^{43}$~erg\,s$^{-1}$); (2) $4.50\pm{0.81}$\% for a limiting
$L_X = 10^{41}$~erg\,s$^{-1}$; and (3) $1.19\pm{0.11}$\% for a
limiting $L_X = 10^{42}$~erg\,s$^{-1}$.  These compare to cluster
active fractions of 6.0\%, 4.9\%, and 1.0\%, respectively. This
excellent agreement between the field and cluster fractions is also
borne out for a brighter magnitude cut. For $M_R < -21.3$ and a
limiting $L_X = 10^{41}$~erg\,s$^{-1}$, we find a fraction of
$12.38\pm{2.26}$\%, which agrees with the \cite{Martini07} value of
9.8\% within 1.5 $\sigma$. For these same magnitude and X-ray
luminosity limits, we also find a fraction of $10.42\pm{4.29}$\% in
the narrower redshift range ($0.01 < z < 0.05$) used by \cite{Sun07},
who find a cluster AGN fraction of 5.5\%. The striking concordance
between cluster and field fractions out to $z \sim 0.3$ implies that
AGN activity at low-redshift does not depend strongly on local
density.

There are various scenarios for why the field and cluster fractions
might be the same. As described in \cite{Martini09}, it is possible
that blue, late-type galaxies in clusters are still in-falling
and have not yet been stripped of their gas by interactions with the
cluster potential. As a result, these galaxies might retain enough cold gas
to fuel an AGN, as is the case in the field. In another scenario, red,
early-type galaxies may keep their cold gas (despite interactions with
cluster potential) and thus continue to host an active SMBH. A third
possibility invokes evolution of the AGN fraction in both clusters and
in the field --- here, AGN activity was more dependent on environment
in the past, \eg, closer to the peak in the AGN number density at
$z \sim 2$, but at the present day galaxies in both environments are
relatively quiescent and the differences between the field and
clusters is not discernible. This last scenario can be tested by
probing the AGN fraction in both clusters and the field at high
redshifts (and possibly in regimes where low-luminosity AGN dominate). 

As discussed in \S \ref{agn_frac} and \S \ref{field_frac}, there is
some evidence that the AGN fraction evolves as a function of
redshift. Though we find the AGN fraction in clusters and in the field
in close agreement at low-redshift, they may begin to differ as we
probe to higher redshift regimes. Employing a method similar to
\cite{Page01}, \cite{Martini09} estimate the field AGN fraction based
on galaxy luminosity functions and compare it to their \Chandra\
cluster fraction (see their Fig. 3). At $z \sim 0.15$, they find the
two in agreement, but at $z \sim 0.8$ the field fraction is nearly a
factor of 5 higher than the cluster fraction. This gives some credence
to a scenario wherein the cluster and field AGN fractions were more
disparate at earlier times.

Other authors \citep{Georgakakis08b,Silverman09a} have investigated
the AGN fraction in groups of galaxies (or as a function of local
over-densities). Using DEEP2 \Chandra\ observations from AEGIS for
$0.7 < z < 0.9$ and $M_B < -20$, \cite{Georgakakis08b} find group and
field fractions of $4.7\pm{1.6}$\% and $4.5\pm{1.0}$\%,
respectively. This group fraction is about 5 times higher than that
reported by \cite{Martini09} in clusters at similar redshifts. This
discrepancy may arise from intrinsic differences in the group and
cluster environments at redshifts near $z \sim 1$, but more likely it
originates from differences in the X-ray and optical selection
criteria of the surveys; \ie\ \cite{Georgakakis08b} include
lower-X-ray-luminosity sources than do \cite{Martini09} and their host
galaxy magnitude limits differ. Using only the AEGIS data,
\cite{Georgakakis08b} report a higher AGN fraction in groups than in
the field at $z \sim 1$, but attribute this to their finding that AGN
are preferentially found in red, luminous galaxies, which are
themselves more likely to reside in higher density environments. They
note that an exception to this trend may exist for the most
X-ray-luminous AGN, which appear in higher numbers in the field.

%--------------------------FIGURE 16---------------------------
\begin{figure*}[tb]
\includegraphics[width=1.0 \textwidth]{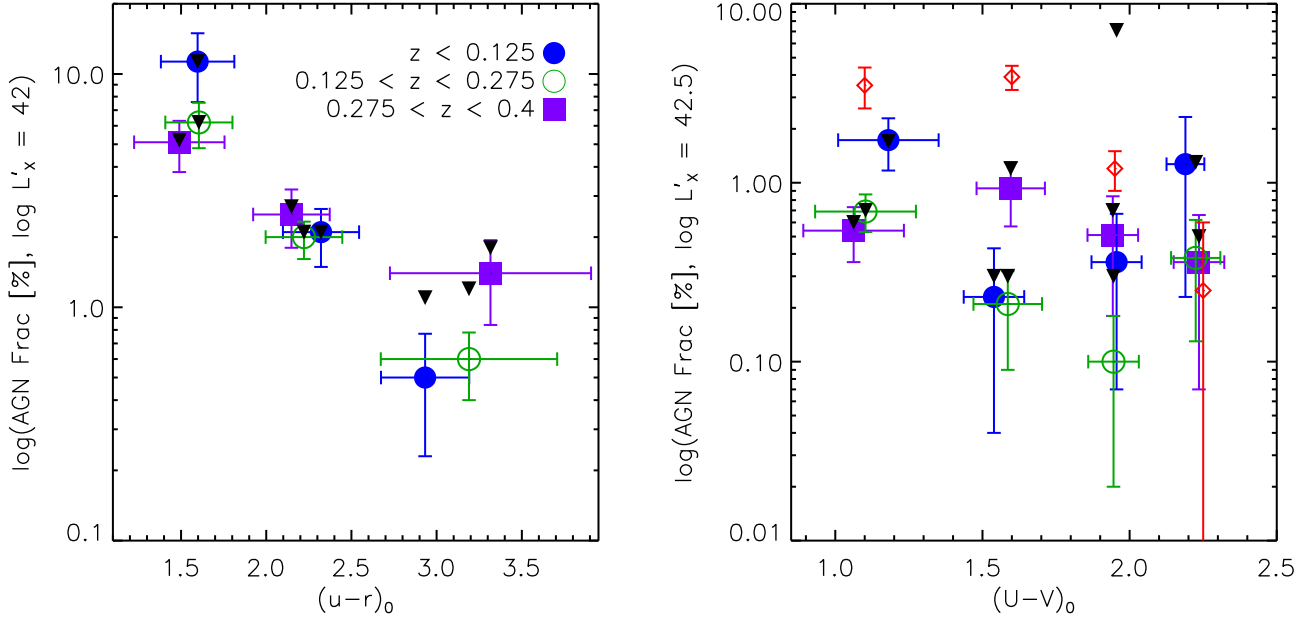}
\caption{({\it Left}) The AGN fraction (log $L_X^{\prime} = 42$, solid
  black line in the right panel of Fig. \ref{mi_lx_z}) for three
  low-redshift samples, binned also by $(u-r)_0$ color (see \S
  \ref{discussion} for discussion). ({\it Right}) Similar to the left
  panel, but the data are binned instead by $(U-V)_0$ color and the
  X-ray luminosity limit is fixed at log $L_X^{\prime} = 42.5$ and
  plotted on a log scale (black dashed line in the right panel of
  Fig. \ref{mi_lx_z}) to compare to zCOSMOS results (open red
  diamonds). Upper limits on the fractions are plotted as solid,
  downward-facing black triangles in both panels (see \S
  \ref{bias}). The trend toward lower AGN fractions at redder colors
  is clearly visible for both color regimes. \\
  (A color version of this figure is available in the online journal.)}
\label{color_trend}
\end{figure*}
%-------------------------------------------------------------

\cite{Silverman09a} also study the X-ray AGN fraction for low-mass
host galaxies ($10.4 < {\rm log}~M_{\star} < 11$) in the field and in
groups, and find no statistical difference between the fractions in
these environments. They do report a trend for log $M_{\star} > 11$,
wherein high X-ray luminosity AGN are less common in more dense
environs, see also \cite{Kauffmann04} and \cite{Constantin08}. For
$0.58 < z < 1.05$ and log $M_{\star} > 10.4$, \cite{Silverman09a} find
a field fraction of $\sim 3.7 \pm 0.5$\% and a group fraction of $\sim
4 \pm 1.2$\%, consistent with the findings of \cite{Georgakakis08b}.

We look to our highest redshift bins to compare our field X-ray AGN
fractions to those of \cite{Georgakakis08b} and
\cite{Silverman09a}. For $0.55 \le z < 0.7$, our only complete mass
bin is $11.6 \le {\rm log}~M_{\star} < 13$, for which we calculate an
upper limit on the AGN fraction of $25.00\pm{20.80}$\%. Note, however,
that we base this estimate of the fraction on only two X-ray-observed
host galaxies and zero X-ray detections and thus do not offer a
sensitive probe of the field fraction (particularly within this
redshift range). For $0.4 \le z < 0.55$, we have two complete mass
bins, $11 \le {\rm log}~M_{\star} < 11.6$ and $11.6 \le {\rm
log}~M_{\star} < 13$, in which we measure AGN fractions of
$5.33\pm{1.70}$\% and $3.70\pm{3.13}$\%. Though these fractions are
broadly consistent with the AEGIS and zCOSMOS studies, they sample
higher mass hosts at lower redshifts --- our median redshifts are
$\langle z \rangle = 0.44$ and $\langle z \rangle = 0.45$ for the low-
and high-mass bins, respectively. To include lower-mass hosts (\eg,
log $M_{\star} > 10.4$), we must set our redshift limit to $z = 0.4$
to achieve a complete sample, and questions about evolution begin to
impede comparisons with the deep field results.

We point out these higher-redshift results to illustrate that small
number statistics continue to stymie measurements of the X-ray AGN
fraction in all environments (field, group, and cluster), particularly
at the faint end of the galaxy population and at high-redshift.

\subsection{AGN Host Galaxy Properties}
\label{host_color}

To investigate the properties of our host galaxies, we show the AGN
fraction in bins of \bestz\ vs. restframe $(u-r)_0$ and $(U-V)_0$
color (Fig. \ref{color_trend}, left and right panels,
respectively). The $(u-r)_0$ blue, green, and red color bins are
$(u-r)_0 \le 1.8$, $1.8 < (u-r)_0 < 2.6$, and $(u-r)_0 \ge 2.6$. The
green bin straddles the $(u-r)_0 = 2.2$ divide between blue and red
galaxies defined by \cite{Strateva01}. The $(U-V)_0$ color bins are
$0.8 < (U-V)_0 < 1.4$, $1.4 < (U-V)_0 < 1.8$, $1.8 < (U-V)_0 < 2.1$,
and $2.1 < (U-V)_0 < 2.4$, selected to match the zCOSMOS color bins
\citep[][their Figure 12b]{Silverman09b}. We apply a mass cut of log
$M_{\star} > 10.6$ (black solid line in the right panel of
Fig. \ref{mass_z}), again to match the zCOSMOS study of host galaxy
colors, and include only the redshift ranges associated with our
samples 1*-3*, as these are complete for this mass cut. For the
$(u-r)_0$ comparison we employ the X-ray limit used throughout this
work ($L_X = 10^{42}$~erg\,s$^{-1}$); for the zCOSMOS $(U-V)_0$
comparison we use $L_X = 10^{42.5}$~erg\,s$^{-1}$ as in
\cite{Silverman09b}.

In the $(u-r)_0$ bins, the fraction of AGN in the blue sequence is
larger than that in either the red sequence or the green valley, for
all three redshift intervals (Fig. \ref{color_trend}, left panel,
colored symbols). This trend toward a larger AGN fraction in the blue
sequence persists when the red point sources, which might bias the red
sequence and/or the green valley toward a lower AGN fraction, are
removed (Fig. \ref{kcorr}, green points; see also \S
\ref{bias}). ``Upper limit'' AGN fractions, calculated without these
red point sources, are shown in Figs. \ref{trends} and
\ref{color_trend} (filled, black, downward-facing triangles). This
result is in contrast to multiple studies, which have found that AGN
lie preferentially in the valley between the red and blue galaxy
sequences, \eg,
\cite{Georgakakis08b,Georgakakis08a,Silverman08a,Schawinski10}; though
\cite{Silverman08a} associate the green valley peak with large-scale
structures at particular redshifts. Our finding agrees, however, with
more recent studies of the zCOSMOS fields
\citep{Silverman09a,Silverman09b}; these authors argue that the peak
reported in the transition region between blue and red galaxies is an
artifact that can be eliminated by applying an appropriate mass cut,
as we have done here.

%--------------------------FIGURE 17---------------------------
\begin{figure*}[tb]
\includegraphics[width=0.5 \textwidth]{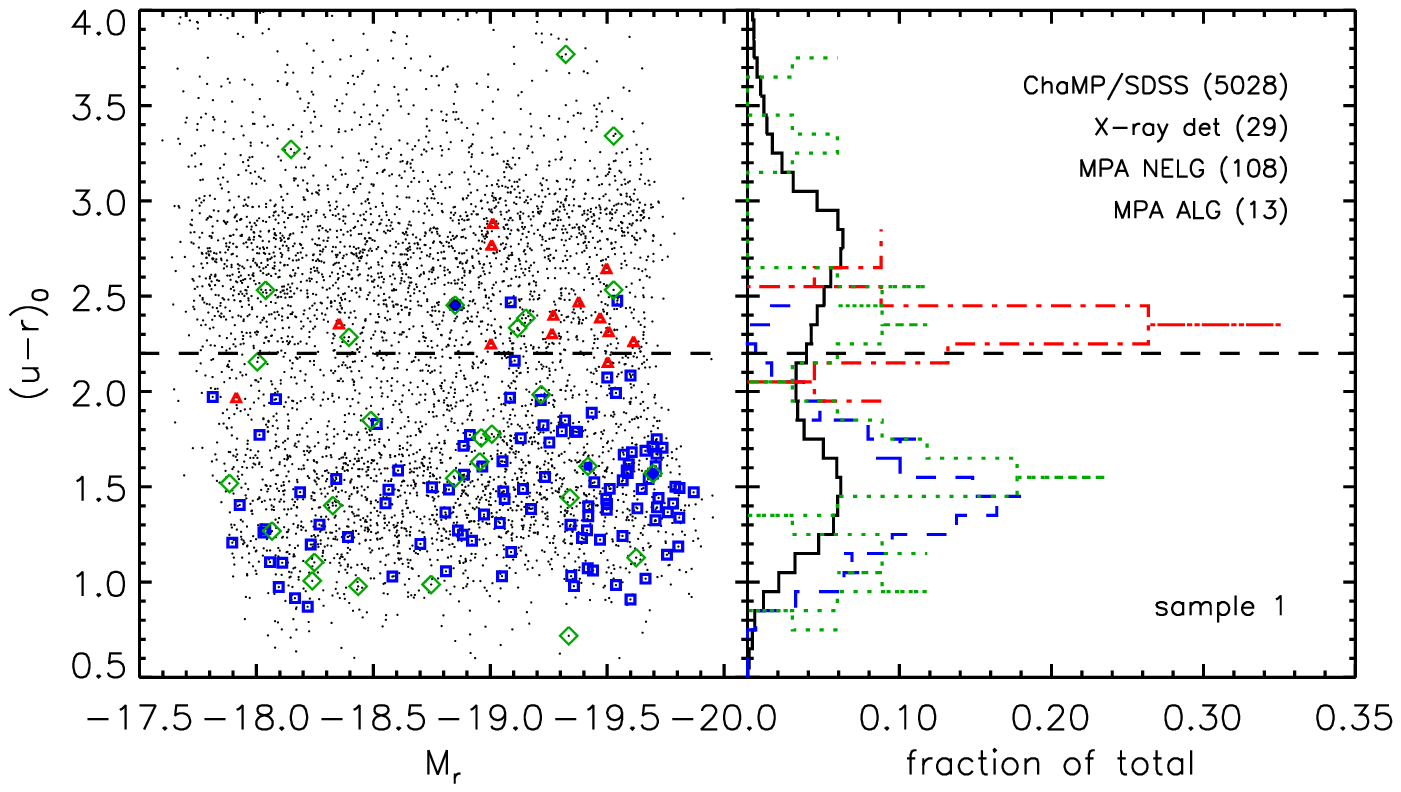}
\includegraphics[width=0.5 \textwidth]{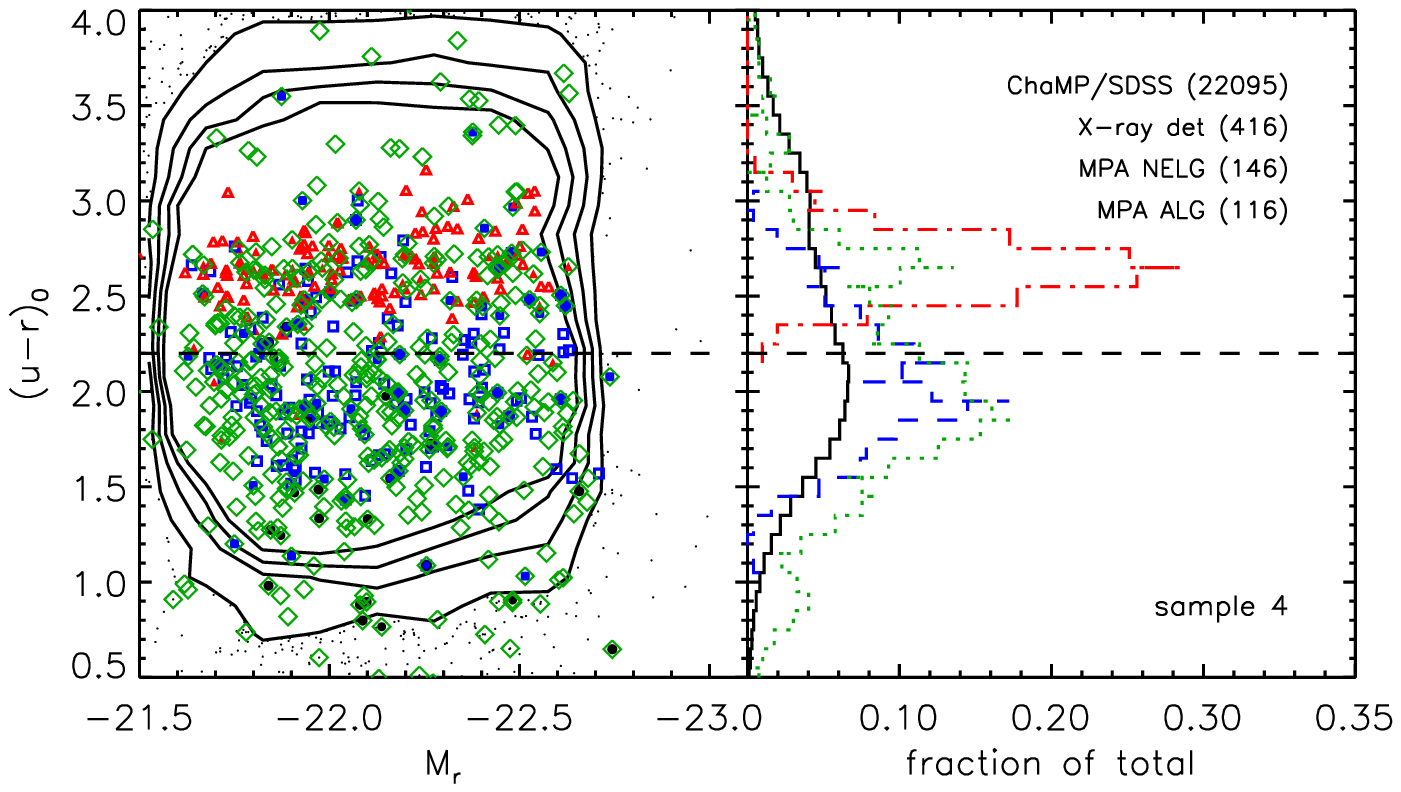}
\includegraphics[width=0.5 \textwidth]{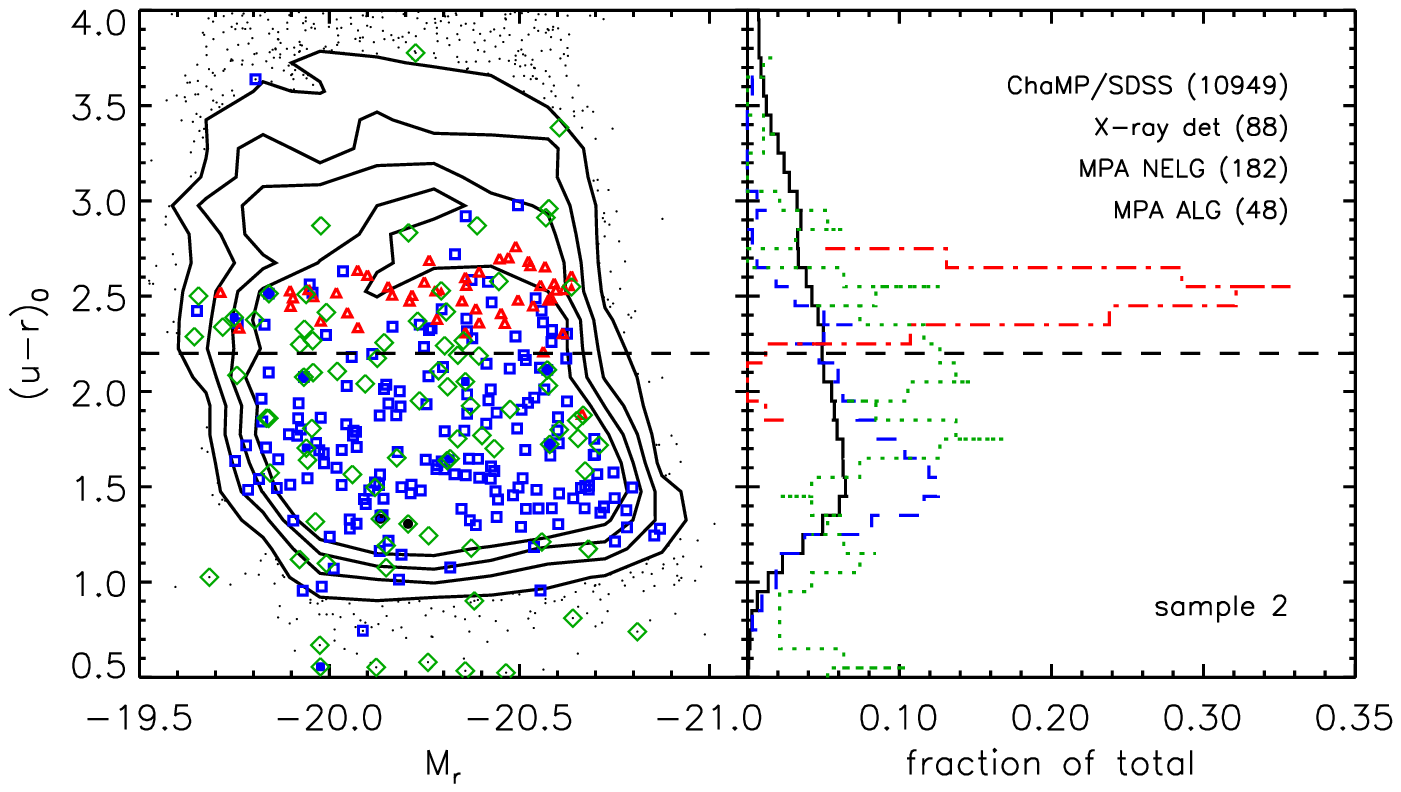}
\includegraphics[width=0.5 \textwidth]{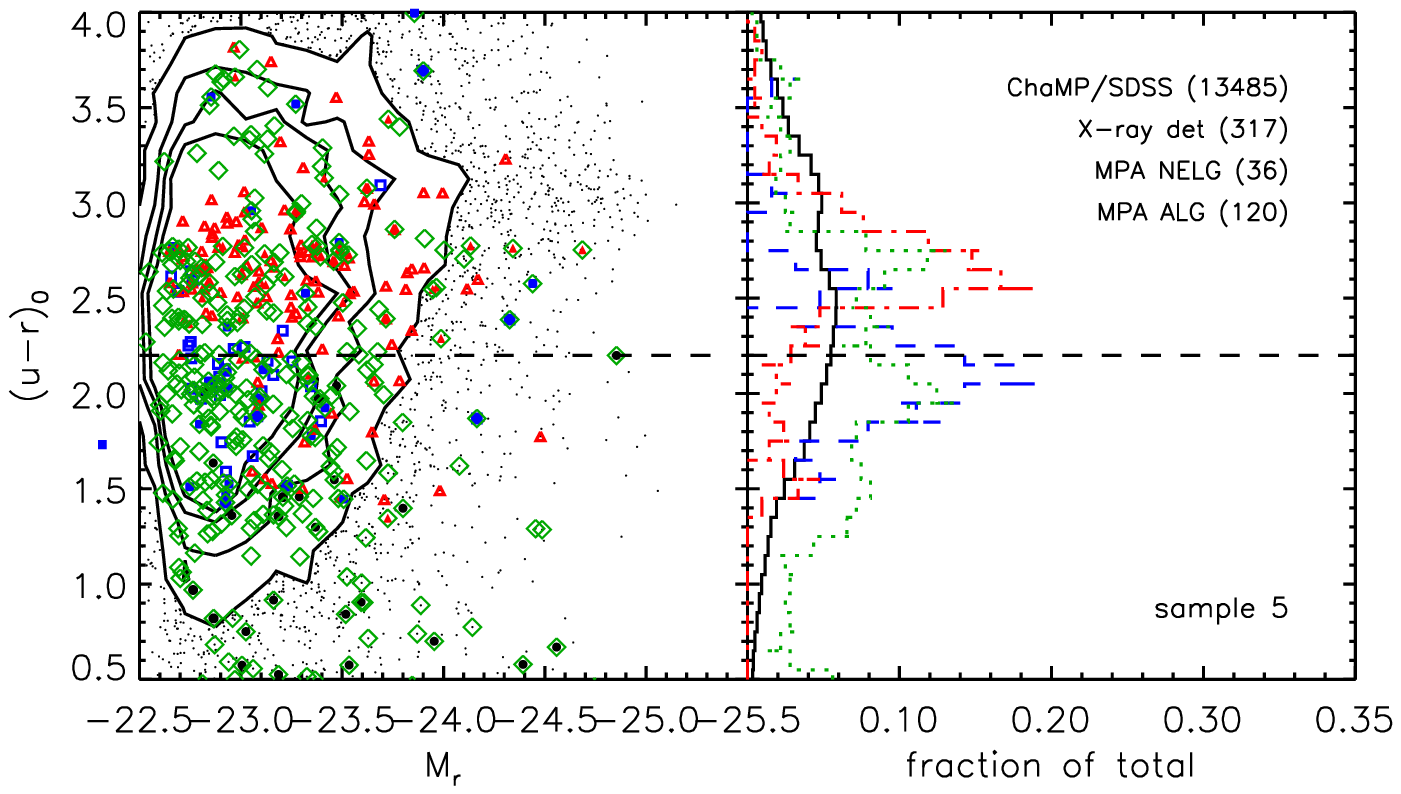}
\includegraphics[width=0.5 \textwidth]{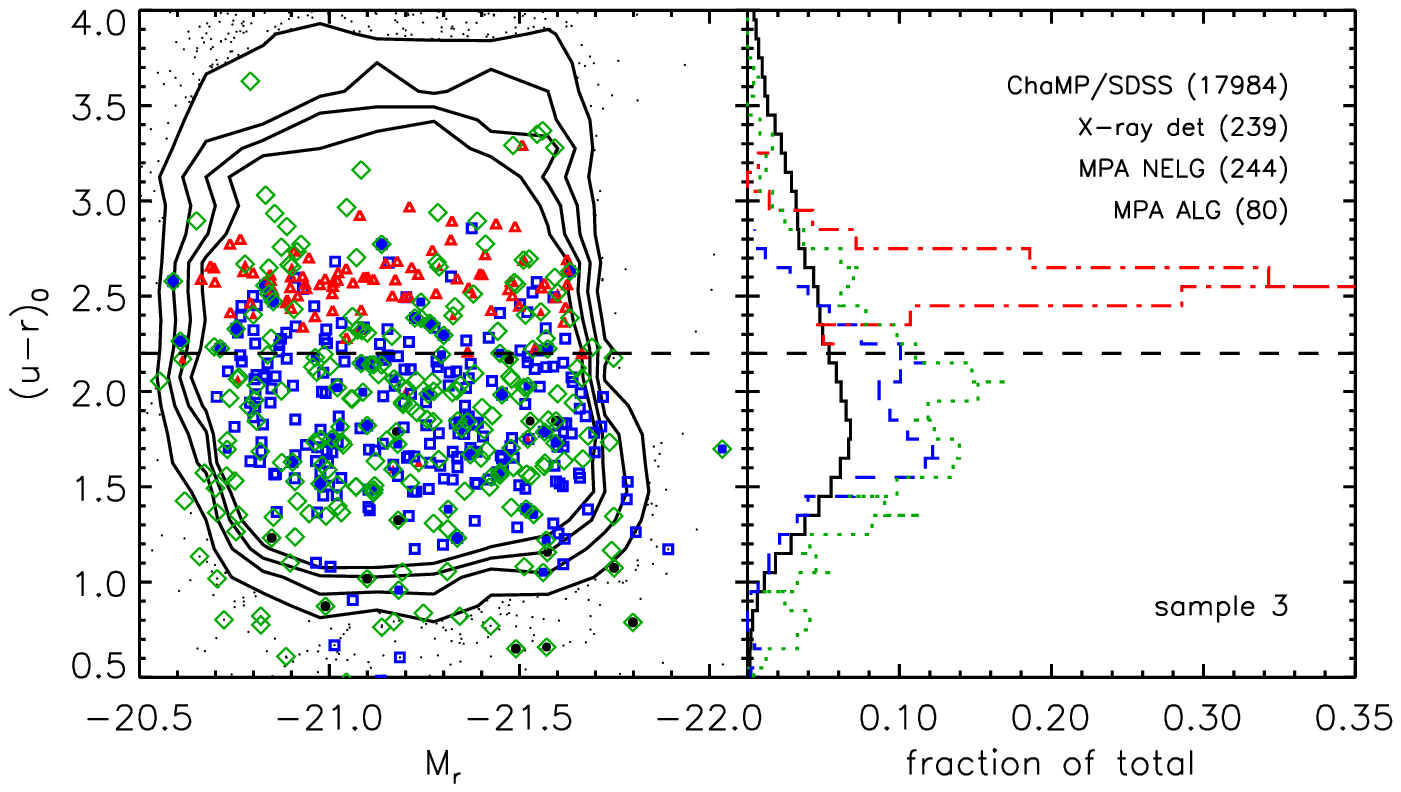}
\caption{({\it Left sub-panels}) Restframe $(u-r)$ as a function of
$r$-band absolute magnitude for the Full galaxy sample (black dots),
the X-ray-detected sample (open green diamonds), and objects with
spectra: ChaMP/SDSS NELG (open blue squares), ChaMP/SDSS
X-ray-detected NELG (filled blue squares), ChaMP/SDSS ALG (open red
triangles), ChaMP/SDSS X-ray-detected ALG (filled red triangles), SDSS
BLAGN (open black circles), and ChaMP BLAGN (filled black
circles). ({\it Right sub-panels}) Histogram of $(u-r_0)$ for the Full
photometric sample (black solid line), the X-ray-detected sample
(green dotted line), the SDSS spectroscopic NELG sample (dashed blue
line), and the SDSS spectroscopic ALG sample (dot-dashed red
line). The $(u-r)_0 = 2.2$ division between the red and blue sequences
from \cite{Strateva01} is shown in both panels (dashed black line). \\
(A color version of this figure is available in the online journal.)}
\label{greenvalley} 
\end{figure*}
%-------------------------------------------------------------

As discussed in \S \ref{bias}, the AGN central engine may contribute
considerable light to the SDSS SED, making the object appear
artificially blue. AGN contamination of the host galaxy light would
result in an overestimate of the number of galaxies along the blue
sequence. We do not expect this effect to be large, however, since the
majority of the X-ray sources in our three lowest redshift bins ($z
\le 0.4$) have $L_X < 10^{44}$~erg\,s$^{-1}$ (Fig. \ref{mi_lx_z}, left
panel); hence, accretion onto the central SMBH is probably not
powerful enough to significantly contaminate the host galaxy light
\citep{Prieto10}. A population of starburst galaxies, whose X-ray
luminosities can be as high as $L_X \sim 10^{42}$~erg\,s$^{-1}$, might
also contaminate our blue sequence, but such X-ray-bright star-forming
galaxies are more likely at higher redshift \citep[$z>0.4$;][]{Yan10}.

To further explore the properties of our sample as a function of host
galaxy properties, we compare our primarily photometric Full galaxy
sample to our spectroscopic sub-sample. Figure \ref{greenvalley} shows
$(u-r)_0$ color vs. absolute $r$-band magnitude ($M_r$; left panels),
and $(u-r)_0$ color histograms (right panels) for each of our five
volume-limited samples. In the left-hand panels we overplot
spectroscopic types for all galaxies that have secure line
identifications in the MPA/JHU DR7 catalog, as well as those
identified via ChaMP spectroscopic follow-up (filled symbols indicate
X-ray detections, open symbols mark X-ray non-detections): blue
squares are narrow emission line galaxies (NELG); red triangles show
absorption line galaxies (ALG); black circles mark broad-line AGN. We
also indicate X-ray-detected sources, with and without spectroscopic
classifications, as open green diamonds. In the right-hand panels, we
show $(u-r)_0$ histograms for the Full galaxy sample (solid black
line), the X-ray-detected sub-sample (green dotted line), as well as
for the MPA/JHU NELG (blue dashed line) and ALG (red dot-dashed line)
samples. The histograms are each normalized to the total number of
objects within the (sub-)sample, shown in parentheses in the relevant
figure.

We see a clear division between the NELG and ALG samples in Fig.
\ref{greenvalley}, with NELGs falling along the blue sequence and ALGs
along the red sequence. We do not see an over-density of X-ray sources
in the green valley between the red and blue sequences, as has
sometimes been reported elsewhere
\citep{Georgakakis08b,Georgakakis08a,Silverman08a,Schawinski10}. In
the color histograms, the Full galaxy sample demonstrates a clear
bi-modality at low redshift, but this trend becomes washed out (or
perhaps shifted) at higher-redshift. This may result from evolution in
the galaxy population, from smearing due to photo-$z$ uncertainties,
or from stellar interlopers. As an illustration of this last
possibility, we note that for Sample 1 (Fig.  \ref{greenvalley}, first
panel) the peak of the red sequence in the Full galaxy sample (solid
black line) is redder than the spectroscopic ALG sample (red
dot-dashed line) --- this offset may be due to a population of M-dwarf
stellar contaminants amongst the objects with photometric
redshifts. Here again, contamination of the host color by a luminous
central engine may force galaxies out of the green or red portions of
the diagram and onto the blue sequence (see \S \ref{bias}), blurring
the expected bi-modality in the underlying galaxy population. However,
since such contamination is unlikely in the majority of the
non-X-ray-detected (Full galaxy) sample, we do not expect the effect
to be large enough to explain the trends we see here.

We also compare the $(U-V)_0$ results (Fig. \ref{color_trend}, right
panel) to the field fraction determined in the 10k catalog of the
zCOSMOS spectroscopic redshift survey
\citep{Silverman09a,Silverman09b}. The \cite{Silverman09b} fractions
for these four $(U-V)_0$ bins (estimated from their Fig. 12b) are
approximately $3.5\pm{0.9}$\%, $3.9\pm{0.6}$\%, $1.2\pm{0.3}$\%, and
$0.25\pm{0.35}$\% (Fig. \ref{color_trend}, right panel, red open
diamonds). We compare these to our $(U-V)_0$ fractions (taking a
weighted averaged over our three redshift bins) of $0.67\pm{0.12}$\%,
$0.26\pm{0.10}$\%, $0.14\pm{0.08}$\%, and $0.40\pm{0.19}$\%. As was
the case for the global field fraction (\S \ref{field_frac}), our
ChaMP/SDSS fractions fall significantly lower than the zCOSMOS
fractions, except in the reddest color bin. The zCOSMOS sample is
almost certainly weighted toward higher redshifts (their redshift
range is $0.1 < z < 1.02$, vs. ours for this comparison, $0.0025 <
z < 0.4$). Thus, we might rather compare only to our AGN fraction
in the highest redshift bin ($0.275 < z < 0.4$), \ie\
$0.54\pm{0.19}$\%, $0.93\pm{0.37}$\%, $0.51\pm{0.33}$\%, and
$0.36\pm{0.3}$\%. Even here, our field AGN fractions are
systematically lower than those found in the zCOSMOS survey; they
disagree by approximately $3~\sigma$ in the blue and green sequences,
by roughly $2~\sigma$ in the red sequence, and agree within the errors
only in the reddest color bin. Despite these differences, both our AGN
fractions and the zCOSMOS fractions are largest for host galaxies with
blue colors. \cite{Silverman09b} propose that this trend might
establish a connection between accretion onto SMBH and star formation
out to $z \sim 1$. Our findings lend additional support to this
hypothesis.
 
It is plausible that the difference between the field AGN fractions
from ChaMP/SDSS and zCOSMOS arises from evolution in the AGN
fraction. If we assume a mean redshift for the zCOSMOS sample of $z
\sim 0.6$, and a mean redshift of $z \sim 0.3$ for our sample 3, we
would expect PLE $(1+z)^3$ evolution in the AGN XLF to yield a zCOSMOS
fraction that is a factor of $\sim 2$ higher than the ChaMP/SDSS
fraction. Such a correction would bring our results closer to the
zCOSMOS blue and green sequence fractions, and achieve agreement in
the red sequence. Note that the fractions in the reddest bin would,
however, begin to disagree. Evolution of the underlying host galaxies
colors may explain these differences, \eg, AGN activity might occur
most often in blue galaxies at higher redshifts ($z > 0.5$), but
increasingly shift toward red galaxies in the local Universe
\citep[see also discussions in][]{Silverman08a,Silverman09b}.

\section{Conclusions}
\label{conclusions}

We present a study of the X-ray-active fraction of field galaxies in
an extensive catalog of more than $100,000$ galaxies with X-ray and
optical coverage from \Chandra\ and SDSS (\about$1,600$ are
X-ray-detected). We combine ChaMP/SDSS spectroscopic and photometric
redshifts with X-ray and optical fluxes (or flux limits) to assign
absolute magnitudes, X-ray luminosities (or limits), masses, and
colors to each galaxy in the sample. With these data we explore the
AGN fraction ($L_X~(0.5-8 {\rm keV}) > 10^{42}$~erg\,s$^{-1}$) in five
independent samples, complete in redshift and $i$-band absolute
magnitude:

{\small
\begin{itemize}
\item $\mathbf{F_{AGN,1}} = 0.16\pm{0.06}$\% ($z \le 0.125$, $-18 > M_i > -20$)
\item $\mathbf{F_{AGN,2}} = 0.50\pm{0.11}$\% ($z \le 0.275$, $-20 > M_i > -21$)
\item $\mathbf{F_{AGN,3}} = 1.27\pm{0.18}$\% ($z \le 0.400$, $-21 > M_i > -22$)
\item $\mathbf{F_{AGN,4}} = 2.85\pm{0.39}$\% ($z \le 0.550$, $-22 > M_i > -23$)
\item $\mathbf{F_{AGN,5}} = 3.80\pm{0.92}$\% ($z \le 0.700$, $M_i < -23$)
\end{itemize}
}

{\noindent The low-redshift bin edge for these fractions is $z =
0.0025$, selected to avoid galactic objects and photometric redshift
artifacts. }

Our robust analysis is enabled by ChaMP's comprehensive sensitivity
maps for ACIS imaging, which allow recognition of
imaged-but-undetected objects, counts limits for 50\% and 90\%
detection completeness, and corresponding flux upper limits at any sky
position. The AGN (or X-ray-active) fraction for each volume-limited
sample, as well as the asymmetric error bars, are calculated via the
$\beta$ distribution --- an ideal statistical tool for evaluating
fractions, particularly when the numerator and denominator are not
drawn from independent samples.

We find excellent agreement between our ChaMP/SDSS field AGN fraction
and the \Chandra\ cluster AGN fraction, for samples restricted to
similar redshift and absolute magnitude ranges. For ChaMP/SDSS field
galaxies with $0.05 < z < 0.31$ and absolute $R$-band magnitude more
luminous than -20, $F_{AGN} = 1.19\pm{0.11}$\%. The \cite{Martini07}
X-ray cluster fraction for this redshift and $M_R$ is $1.0$\%. We find
good agreement between our field fraction and the cluster fraction for
a variety of other absolute $R$-band magnitude and X-ray luminosity
limits as well. Our results are also broadly consistent with measures
of the field AGN fraction in narrow, deep fields, though differences
in the optical selection criteria, redshift coverage, and possible
cosmic variance between fields introduce larger uncertainties in these
comparisons.

In our analysis of the AGN fraction, as well as in our comparison to
deep field studies, we find evidence that the AGN fraction evolves
with redshift. Our data are consistent with either $(1+z)^3$ or
$(1+z)^4$ evolution, \ie\ the two most common fits to the X-ray AGN
luminosity function, but likely also depend on evolution in the host
galaxy population. These findings are tantalizing, but poorly
constrained and require high-quality wide-field data at redshifts out
to (and above) $z \sim 1$ for verification. We test the impact of host
galaxy color (for stellar masses log $M_{\star} > 10.6$) and find that
the AGN fraction is largest for hosts with blue colors, though this
trend too may depend on redshift. Since galaxy and AGN evolution are
both affected by environment, we are embarking on a clustering study
of AGN activity vs. environment to quantify these properties. A
comparison between cluster, group, and field fractions at redshifts
approaching $z = 2$ would be particularly compelling.

\acknowledgments

We thank the referee for comments that improved this manuscript;
Nicolas B. Cowan for useful discussions about the $\beta$ distribution
and Monte Carlo simulations; Adam D. Myers for providing QSO redshift
probability distributions for the Monte Carlo simulations; and John
D. Silverman for assistance with the comparisons to the zCOSMOS AGN
fractions. Support for this work was provided by the National
Aeronautics and Space Administration through \Chandra\, Award Number
AR7-8015A-R0, AR9-0020X, and GO0-11129B issued by the \Chandra\, X-ray
Observatory Center, which is operated by the Smithsonian Astrophysical
Observatory for and on behalf of the National Aeronautics Space
Administration under contract NAS8-03060.  This research was also
supported in part by the National Science Foundation under Grant
No. NSF PHY05-51164. D.H. acknowledges support from the NASA Harriett
G. Jenkins Predoctoral Fellowship Program and the University of
Washington Astronomy Department's Jacobsen Fund.  We acknowledge use
of the NASA/IPAC Extragalactic Database (NED), operated by the Jet
Propulsion Laboratory, California Institute of Technology, under
contract with the National Aeronautics and Space Administration.

Funding for the SDSS and SDSS-II has been provided by the Alfred
P. Sloan Foundation, the Participating Institutions, the National
Science Foundation, the U.S. Department of Energy, the National
Aeronautics and Space Administration, the Japanese Monbukagakusho, the
Max Planck Society, and the Higher Education Funding Council for
England. The SDSS Web Site is http://www.sdss.org/. The SDSS is
managed by the Astrophysical Research Consortium for the Participating
Institutions. The Participating Institutions are the American Museum
of Natural History, Astrophysical Institute Potsdam, University of
Basel, Cambridge University, Case Western Reserve University,
University of Chicago, Drexel University, Fermilab, the Institute for
Advanced Study, the Japan Participation Group, Johns Hopkins
University, the Joint Institute for Nuclear Astrophysics, the Kavli
Institute for Particle Astrophysics and Cosmology, the Korean
Scientist Group, the Chinese Academy of Sciences (LAMOST), Los Alamos
National Laboratory, the Max Planck Institute for Astronomy (MPIA),
the Max Planck Institute for Astrophysics (MPA), New Mexico State
University, Ohio State University, University of Pittsburgh,
University of Portsmouth, Princeton University, the United States
Naval Observatory, and the University of Washington. 

{\em Facilities:} \facility{CXO, Sloan, FLWO:1.5m (FAST spectrograph),
 Magellan:Baade (LDSS2 imaging spectrograph), Magellan:Clay (IMACS),
 WIYN (Hydra)}

\appendix

In \S \ref{xactive_frac} we define the fraction of X-ray-active
galaxies. We apply both optical and X-ray completeness
criteria to insure an accurate measure of the fraction. In this
Appendix, we describe the objects that appear in our volume-limited
optical samples, but do not pass the X-ray cuts. These ``dropped''
objects can be described by:

\begin{equation}
N_{lim,drop} = \sum^{z_{max}}_{z_{min}} \sum^{M_{bright}}_{M_{faint}} \sum^{\infty}_{L_X^{\prime}} G(L_{X,lim},M_i,bestz)~,
\end{equation}

\begin{equation}
N_{det,drop} = \sum^{z_{max}}_{z_{min}} \sum^{M_{bright}}_{M_{faint}} \sum^{\infty}_{L_X^{\prime}} \sum^{\infty}_{L_X^{\prime}} X(L_X,L_{X,lim},M_i,bestz)~,
\end{equation}

\noindent{where $N_{lim,drop}$ quantifies the number of X-ray-observed
objects ``dropped'' from the denominator, and $N_{det,drop}$ describes
the number of X-ray-detected sources eliminated from the numerator. }

%---------------------------FIGURE A1-------------------------
\begin{figure}[b]
\includegraphics[width=0.5 \textwidth]{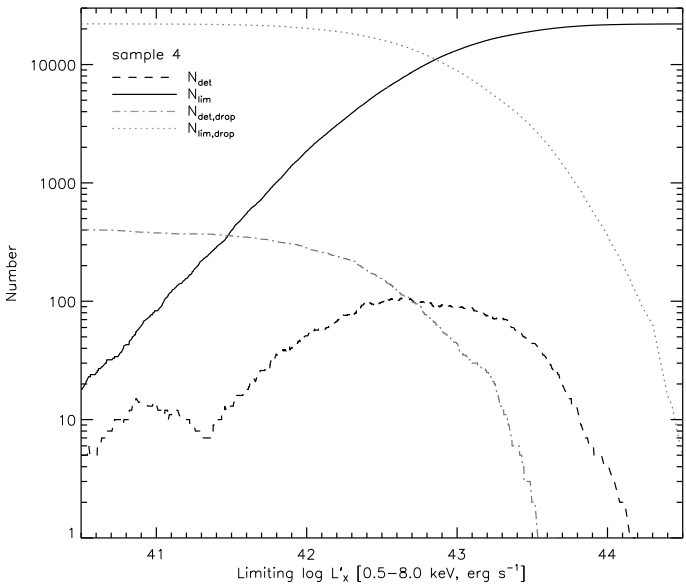}
\caption{The numbers of X-ray-detected ($N_{det}$) and X-ray-observed
  ($N_{lim}$) galaxies included in the X-ray-active fraction at each
  X-ray threshold for sample 4 are shown as black dashed and solid
  lines, respectively. The numbers of objects that are ``dropped'' at
  each limit are also shown (gray dot-dashed line for X-ray-detections
  [$N_{det,drop}$], gray dotted line for X-ray-observed objects
  [$N_{lim,drop}$]). This plot makes explicit the shape of the X-ray
  active fraction curve in Fig. \ref{agnfrac_fig}, \ie\ while the
  number of X-ray-detections included in the sample ranges from a few
  to several tens of objects between log $L_X$ of 40.5 and 44, the
  number of X-ray-observed galaxies increases from about 20 to more
  than 20,000 in the same range. In this optically-complete sample the
  majority of the brightest X-ray sources make it into the fraction,
  while many X-ray-low-luminosity sources are eliminated.}
\label{agnfrac_num}
\end{figure}
%-------------------------------------------------------------

In Fig. \ref{agnfrac_num}, we show the behavior of
$N_{det},~N_{lim},~N_{det,drop},~{\rm and}~N_{lim,drop}$ as a function
of limiting X-ray luminosity for sample 4. Comparing this to the top,
right panel of Fig. \ref{agnfrac_fig}, it is clear that the high
X-ray-active fraction below log $L_X^{\prime} \sim 41.4$ can be
explained by a relatively small number of both X-ray-detected and
X-ray-observed galaxies. At higher limiting X-ray luminosities, the
number of X-ray detections in the numerator increases (until log
$L_X^{\prime} \sim 43$), but the number of galaxies that might have
been detected at this threshold X-ray luminosity (the denominator)
increases even more rapidly, leading to an overall decline in the
X-ray-active fraction. At the highest luminosities the number of
X-ray-detected sources falls off sharply while the number of
X-ray-observed galaxies continues to climb; the fraction drops to
zero, accordingly. The error intervals shown in Fig. \ref{agnfrac_fig}
are intuitive insofar as they are largest where the number of objects
used to determine the fraction is small, and smallest in the opposite
extreme.

\end{document}